\documentclass[11pt,a4paper]{amsart}
\usepackage{amsmath,amssymb, amsbsy}
\usepackage{subfigure}
\usepackage{graphpap,latexsym,epsf}
\usepackage{color,psfrag}
\usepackage[dvips]{graphicx}
\usepackage[hidelinks]{hyperref}
\usepackage{enumerate}
\usepackage{bbm}
\usepackage{relsize}
\renewcommand{\leq}{\leqslant}

\begin{document}
\newcommand{\dyle}{\displaystyle}
\newcommand{\R}{{\mathbb{R}}}
\newcommand{\Hi}{{\mathbb H}}
\newcommand{\Ss}{{\mathbb S}}
\newcommand{\N}{{\mathbb N}}
\newcommand{\Rn}{{\mathbb{R}^n}}
\newcommand{\F}{{\mathcal F}}
\newcommand{\ieq}{\begin{equation}}
\newcommand{\eeq}{\end{equation}}
\newcommand{\ieqa}{\begin{eqnarray}}
\newcommand{\eeqa}{\end{eqnarray}}
\newcommand{\ieqas}{\begin{eqnarray*}}
\newcommand{\eeqas}{\end{eqnarray*}}
\newcommand{\f}{\hat{f}}
\newcommand{\Bo}{\put(260,0){\rule{2mm}{2mm}}\\}
\newcommand{\1}{\mathlarger{\mathlarger{\mathbbm{1}}}}

%- Theorems and similar stuff: ------

\theoremstyle{plain}
\newtheorem{theorem}{Theorem} [section]
\newtheorem{corollary}[theorem]{Corollary}
\newtheorem{lemma}[theorem]{Lemma}
\newtheorem{proposition}[theorem]{Proposition}
\def\neweq#1{\begin{equation}\label{#1}}
\def\endeq{\end{equation}}
\def\eq#1{(\ref{#1})}

%- Definitions -----------------------------

\theoremstyle{definition}
\newtheorem{definition}[theorem]{Definition}
\newtheorem{remark}[theorem]{Remark}
\numberwithin{figure}{section}

\title[It\^o vs H\"anggi--Klimontovich]{It\^o versus H\"anggi--Klimontovich}

\author[C. Escudero]{Carlos Escudero}
\author[H. Rojas]{Helder Rojas}
\address{}
\email{}


\date{\today}

\begin{abstract}
Interpreting the noise in a stochastic differential equation, in particular the Itô versus Stratonovich dilemma, is a problem that has generated
a lot of debate in the physical literature. In the last decades, a third interpretation of noise, given by the so-called H\"anggi--Klimontovich
integral, has been proposed as better adapted to describe certain physical systems, particularly in statistical mechanics. Herein, we introduce
this integral in a precise mathematical manner and analyze its properties, signaling those that have made it appealing within the realm of physics.
Subsequently, we employ this integral to model some statistical mechanical systems, such as the random dispersal of Langevin particles and the
relativistic Brownian motion. We show that, for these classical examples, the H\"anggi--Klimontovich integral is worse adapted than the
Itô integral and even the Stratonovich one.
\end{abstract}
\maketitle

\section{Introduction}\label{intro}

A very large number of physical systems are mathematically modelled by means of differential equations.
Among the simplest models of a time evolution, one finds the initial value problem for an ordinary differential equation of the type
	\begin{equation}\label{edo_0}
	\frac{\mathrm{d}x_t}{\mathrm{d} t}=f(x_t,t), \quad t\geq 0, \quad \left. x_t \right|_{t=0}=x_0,
\end{equation}
where $f:\mathbb{R}\times\mathbb{R}^{+}\longrightarrow \mathbb{R}$ is a given function and $x_0$ is a given real number. 
It has long been recognized that many physical systems are affected by random perturbations, which may be of different origins, and that affect
the system dynamics in a non-negligible manner \cite{Gardiner,HL84,vankampen}. In order to incorporate such random perturbations in model \eqref{edo_0},
it is customary to modify this equation with the introduction of a ``white noise'', so the resulting model reads
\begin{equation}\label{eqf}
	\frac{\mathrm{d}X_t}{\mathrm{d} t}=f(X_t, t)+g(X_t, t)\,\xi_t, \quad t\geq 0, \quad \left. X_t \right|_{t=0}=X_0,
\end{equation}
where $g:\mathbb{R}\times\mathbb{R}^{+}\longrightarrow \mathbb{R}$ is a given function and $X_0$ is a given random variable; of course, the solution
$X_t$ has become a stochastic process rather than a deterministic function. The random function $\xi_t$, which is the ``white noise'', presents
serious mathematical difficulties. While it is a well-defined stochastic process in discrete time, it is not so in continuous time \cite{evans}, unless one
considers it as a distribution-valued stochastic process \cite{holden}. There is, however, a well-known way to circumvent stochastic distributions, which
is the classical theory of stochastic differential equations (SDEs). To cast equation \eqref{eqf} in the form of a well-defined SDE one should, first
of all, consider its integral version
\begin{equation}\label{wn_interpretation}
    	X_t=X_0+\int\limits_0^t f(X_s, s)\,\mathrm{d}s+ ``\int\limits_0^t g(X_s, s)\, \xi_s \, \mathrm{d}s\,\,",\quad t\geq 0.
\end{equation}
The second, and fundamental, step is to give a precise meaning to the integral between quotes, which cannot be considered either as a Riemann or
as a Lebesgue integral. This program was first successfully carried out by It\^o, who, by introducing his stochastic integral, was able to translate
this formal model into a well-defined mathematical object \cite{ito1, ito2}:
\begin{equation}\label{Ito_interpretation}
    	X_t=X_0+\int\limits_0^t f(X_s, s)\,\mathrm{d}s+ \int\limits_0^t g(X_s, s)\,\mathrm{d}W_s\,\, ,\quad t\geq 0.
\end{equation}
Note that a key point in this derivation is noting that the formal white noise process is, intuitively, the derivative of Brownian motion: $``\xi_t=dW_t/dt"$;
herein we keep this quoted expression as an informal philosophical idea, but for a precise statement see \cite{holden}.
The peculiarity of It\^o integration is its set of associated calculus rules, which differ from the classical ones that emanate from the Leibniz-Newton differential calculus. Later on, a new stochastic integral able to mimic these classical rules appeared in the literature: the Stratonovich
integral \cite{stratonovich}. If we employ this theory of stochastic integration, equation \eqref{wn_interpretation} becomes
\begin{equation}\label{Strat_interpretation}
    	X_t=X_0+\int\limits_0^t f(X_s, s)\,\mathrm{d}s+ \int\limits_0^t g(X_s, s)\circ\mathrm{d}W_s,\quad t\geq 0,
\end{equation}
where the last integral denotes the Stratonovich one. In fact, one can replace the formal expression
$$
``\int\limits_0^t g(X_s, s)\, \xi_s \, \mathrm{d}s\,\,"
$$
by one out of infinitely many well-defined mathematical objects to obtain a well-posed equation. Despite this infinite multiplicity,
most works in the literature have opted to choose either the It\^o or the Stratonovich integral. To select a concrete meaning for this
quoted integral in a given applied model has been termed as to choose an \emph{interpretation of noise} for that model. While the interpretation
of noise can be Itô, Stratonovich, or something else, as already mentioned, there has been a strong preference towards the first two in the literature,
at least traditionally \cite{Kampen}.

The theory of stochastic differential equations of either Itô or Stratonovich type is well established \cite{evans,kuo,oksendal}. It relies on the
properties of these integrals. The Itô integral can be defined as
\begin{equation}\label{Ito_integral}
    	 \int\limits_0^t g(X_s, s) \mathrm{d}W_s:=\lim_{\|\Delta_n\|\to 0}\sum_{j=1}^{n}g(X_{t_{j-1}},t_j)(W_{t_j}-W_{t_{j-1}})\quad \textrm{in probability},
\end{equation}
where $\Delta_n=\{t_0, t_1, \ldots,t_{n-1}, t_n\}$ is a partition of $[0, t]$, for which $\|\Delta_n\|:=\max_{1\leq j \leq n}(t_j-t_{j-1})$ is its diameter.
This is sometimes referred to as a weak sense of the Itô integration, whereas the strong or classical sense results from substituting the limit
in probability by a limit in mean square. In words, the Itô integral is defined as a suitable limit of Riemann sums in which the integrand is evaluated at the left endpoint of each subinterval. The Stratonovich integral, at least as it is mainly conceived in the physics literature, would be the corresponding
limit of Riemann sums but with the integrand evaluated at the midpoint of each subinterval; to be precise:
\begin{equation}\label{Strat_integral}
    	 \int\limits_0^t g(X_s, s)\circ\mathrm{d}W_s:=\lim_{\|\Delta_n\|\to 0}\sum_{j=1}^{n}g(X_{(t_{j}+t_{j-1})/2},(t_{j}+t_{j-1})/2)(W_{t_j}-W_{t_{j-1}})\quad \textrm{in probability}.
\end{equation}
This definition might be slightly modified in mathematical texts. The mathematical properties of both integrals have been studied \cite{evans,kuo,oksendal}
and their applications in physics have been discussed \cite{bhattacharyay2020generalization,bhattacharyay2025active,Mannella,Kampen}; for their applications beyond physics one can see, for instance, reference \cite{ontology}. Actually, the preference in the use of either of these two integrals
in physical models has generated much debate along the decades, and this twofold choice has been known as the \emph{Itô versus Stratonovich dilemma}.
As a result of this debate, a folkloric thought emerged: the Itô interpretation of noise should be preferred in mathematics while the
Stratonovich interpretation should be preferred in physics. Of course, this simplistic rule of thumb constitutes an erroneous approach to such a complex
question. More detailed discussions on this dilemma can be found in \cite{bhattacharyay2020generalization,bhattacharyay2025active,ce,cescudero,cescudero2,em},
where it is shown how the Itô integral might overtake the Stratonovich one in many physical instances (see also \cite{roberts2025stochastic}).

Nevertheless, this preeminence of the Itô versus Stratonovich dilemma in the literature has not hidden the fact that there is an infinite number
of potential meanings of the quoted integral in equation \eqref{wn_interpretation}. Simply by comparison with the other two integrals and symmetry,
one is tempted to propose a third variant:
\begin{equation}\label{HK_integral}
    	 \int\limits_0^t g(X_s, s)\bullet\mathrm{d}W_s:=\lim_{\|\Delta_n\|\to 0}\sum_{j=1}^{n}g(X_{t_j},t_j)(W_{t_j}-W_{t_{j-1}})\quad \textrm{in probability},
\end{equation}
that is, to use the same scheme, but evaluating the integrand at the right endpoint of each subinterval. Perhaps surprisingly, perhaps not at all,
this integral has also been used in the physical literature, at least at the formal level. It is known as the H\"anggi--Klimontovich integral, after its
introduction in the seminal works by H\"anggi \cite{hanggi2}, H\"anggi and Thomas \cite{hanggi3}, and Klimontovich \cite{klimont1,klimont2,klimont3}.
This introduction, as already mentioned, was formal, since these works focused on its good properties as a modelling tool in the field of
statistical physics. This integral has not passed unnoticed; on the contrary, it has been referenced quite often in the physical literature due
to its good properties as a mathematical descriptor of different magnitudes and systems. Some examples are relativistic Brownian motion
\cite{dunkelt,dunkel,dunkel2} and other instances of nonlinear Brownian motion \cite{lindner}, or, in general, randomly dispersing particles with
position-dependent diffusivities \cite{volpe2,ottinger,lau,lb,Mei,pacheco2025hetero,rgjf,sokolov,tsekov,volpe}, where this list is not meant to be exhaustive.
Some remarks are now in order: first, this integral is not uniquely known as the H\"anggi--Klimontovich one; sometimes authors refer to it as
the backward-Itô, anti-Itô, isothermal, or kinetic integral. We also note that these terms sometimes diverge in meaning when applied to higher dimensions,
but generally agree when restricted to one-dimensional diffusions. Indeed, the higher-dimensional setting presents particularities and interesting questions as discussed in \cite{escudero2025beneath}.
Second, the majority of papers in the physical literature that refer to this integral do so to
acclaim its good properties (like its simple interplay with the fluctuation-dissipation relation), while criticisms are scarce. Finally, all the works
we are aware of treat this integral formally, most of them restricting their interaction with it to deal with its associated Fokker-Planck equation
(assuming its existence without proof), and with limited assessment of its sample paths.

In this work, we try to bridge the gap that the development of the theory of the H\"anggi--Klimontovich integral in the physical literature has left.
To this end, we construct the full mathematical theory of this type of stochastic integration. We use as starting point reference \cite{hanggi1},
where this program is outlined, but the precise mathematical details are left open. In section \ref{One_HKI} we introduce this integral precisely,
in section \ref{multidhki} we introduce its multidimensional generalization, and in section \ref{sechkisdes} we study its associated stochastic
differential equations. We note that, as happens in the case of the Stratonovich integral \cite{stratonovich}, the theory of stochastic differential equations follows from the multidimensional integral. In section \ref{fpeab} we derive the corresponding Fokker-Planck equation, along with the properties of its solution that have made this integral particularly appealing for its application in several fields within physics. In section \ref{secappdiff} we illustrate its use in several physical examples, but with one particularity: we prove that its properties yield physically inconsistent results in all of these cases, while the Itô and even the Stratonovich integrals are able to model these systems correctly. This section therefore complements the large number of studies that have found the good adaptability of the H\"anggi--Klimontovich integral to model physical systems.
In section \ref{secextending} we further extend the definition and compare it with the backward integral introduced by Russo and Vallois. Finally, in
section \ref{conclusions}, we draw our main conclusions.

\section{H\"anggi--Klimontovich Integral}\label{One_HKI}

\subsection{Definition}
We start by integrating functions of the Wiener process, or Brownian motion process, for simplicity.
First, we establish some notation.
Let $(\Omega,\mathcal{F},(\mathcal{F}_t)_{t\geq 0},\mathbb{P})$ be a completed filtered probability space in which a Brownian motion $(W_t)_{t\geq 0}$ is defined. We denote by $\mathbb{E}$ the expectation with respect to $\mathbb{P}$.
Consider any pair of non-negative numbers $a$ and $b$ satisfying $a<b$, and denote by
\[
\mathcal{C}^1(\mathbb{R},\mathbb{R}):=\{ \varphi: \mathbb{R} \longrightarrow \mathbb{R}~|~ \varphi\textrm{ is differentiable and its derivate is continuous}\}.
\]

\begin{definition}[Integration of smooth functions with respect to Brownian motion]\label{def_simple}
For any $\Phi\in \mathcal{C}^1(\mathbb{R},\mathbb{R})$ the H\"anggi--Klimontovich integral of $(\Phi(W_t))_{t\geq 0}$ in the interval $[a,b]$ with respect to the Brownian motion $(W_t)_{t\geq 0}$ is defined as
	\begin{equation}\label{HLI_cs}
	\int\limits_a^b \Phi(W_t)\bullet \mathrm{d}W_t :=\lim_{\|\Delta_n\|\to 0}\sum_{j=1}^{n}\Phi(W_{t_j})(W_{t_j}-W_{t_{j-1}})\quad \textrm{in probability},
	\end{equation}
where $\Delta_n=\{t_0, t_1, \ldots, t_{n-1}, t_n\}$ is a partition of $[a, b]$ and $\|\Delta_n\|:=\max_{1\leq j \leq n}(t_j-t_{j-1})$ is its diameter.
\end{definition}

In the sequel, we show that under the assumptions established in Definition ~\ref{def_simple} the limit on the right-hand side of Equation (\ref{HLI_cs}) exists. Furthermore, we show that this integral is connected to the Itô integral through a simple formula.

\begin{proposition}\label{simprop}
For any $\Phi \in \mathcal{C}^1(\mathbb{R},\mathbb{R})$ the limit~\eqref{HLI_cs} exists and moreover we have that the identity
\begin{equation}\label{eq:ibf}
\int\limits_a^b \Phi(W_t)\bullet \,\mathrm{d}W_t=\int\limits_a^b \Phi(W_t)\,\mathrm{d}W_t+\int\limits_{a}^{b}\Phi^{\prime}(W_t)\,\mathrm{d}t \quad \textrm{holds almost surely}.
\end{equation}
\end{proposition}

\begin{proof}
The proof follows the ideas employed in \cite{stratonovich}. First note that both summands on the right-hand side of Equation \eqref{eq:ibf} are
almost surely well-defined under the stated assumptions, see Theorem 5.3.3 in ~\cite{kuo}. In the sequel, we  prove that the integral on the left-hand side is well-defined and the equality~\eqref{eq:ibf} holds.
Since $\Phi\in \mathcal{C}^1(\mathbb{R},\mathbb{R})$, the It\^o integral is defined as the limit in probability
\begin{equation}\label{II}
\int\limits_a^b \Phi(W_t) \,\mathrm{d}W_t =\lim_{\|\Delta_n\|\to 0}\sum_{j=1}^{n}\Phi(W_{t_{j-1}})(W_{t_j}-W_{t_{j-1}})
\end{equation}
for any partition $\Delta_n$ satisfying
$\|\Delta_n\|\to 0$ as $n\to \infty$.
Now, for such partitions $\Delta_n$ we consider the difference between the limit expressions on the right-hand sides of~\eqref{HLI_cs}
and~\eqref{II} and  find
\begin{equation}\label{eq:diferencia}
\begin{split}
D_{\Delta_n} &= \sum_{j=1}^{n}\Big[\Phi(W_{t_j})-\Phi(W_{t_{j-1}})\Big](W_{t_j}-W_{t_{j-1}}) \\
&= \sum_{j=1}^{n}\Phi^{\prime}((1-\theta_j)W_{t_{j-1}}+\theta_j W_{t_j})(W_{t_j}-W_{t_{j-1}})^2,
\end{split}
\end{equation}
where we have used that $\Phi\in \mathcal{C}^1(\mathbb{R},\mathbb{R})$ along with the Mean Value Theorem,
and $0 < \theta_j < 1$, $j=1,\ldots,n$ are fixed parameters. Then, using Lemma~7.2.1 and Lemma~7.2.3 from \cite{kuo}, it follows that
\begin{equation}\label{eq:limitderi}
\lim_{\|\Delta_n\|\to 0}\sum_{j=1}^{n}\Phi^{\prime}\big((1-\theta_j)W_{t_{j-1}}+\theta_j W_{t_j}\big)(W_{t_j}-W_{t_{j-1}})^2=\int\limits_{a}^{b}\Phi^{\prime}(W_t)\,\mathrm{d}t
\end{equation}
in probability.
By \eqref{II}, \eqref{eq:diferencia}, and \eqref{eq:limitderi} we deduce the existence of the expression on the left-hand side of Equation \eqref{HLI_cs}.
Finally, by taking a subsequence if necessary, we conclude
\begin{equation*}
\int\limits_a^b \Phi(W_t)\bullet \,\mathrm{d}W_t=\int\limits_a^b \Phi(W_t)\,\mathrm{d}W_t+\int\limits_{a}^{b}\Phi^{\prime}(W_t)\,\mathrm{d}t \quad \textrm{almost surely}.
\end{equation*}
\end{proof}

\begin{remark}
Using the language of stochastic differentials, what we will do in the remainder of this work, the equality in the statement of
this proposition can be written in the form:
\begin{equation*}
\Phi(W_t)\bullet \,\mathrm{d}W_t = \Phi(W_t)\,\mathrm{d}W_t + \Phi^{\prime}(W_t)\,\mathrm{d}t.
\end{equation*}\end{remark}
\vspace{0.06mm}

\subsection{Extending the definition}\label{HKI}

In the previous section, we defined the integral with respect to Brownian motion. However, in many applications (notably, stochastic differential equations) it is necessary to consider a wider family of processes, beyond Brownian motion, for which it is possible to extend our definition. To this end we define a subclass of Markov processes with continuous sample paths almost surely, the so-called diffusion processes, of which the Brownian motion is a particular case \cite{kuo, oksendal}.

\begin{definition}[$\mathbb{R}$-valued diffusion processes]\label{definition_diffusion} A Markov process $(X_t)_{a\leq t \leq b}$ is called a {diffusion process} if it satisfies the following three conditions for any $t\in[a,b]$, $x \in \mathbb{R}$, and $\delta>0$:

\begin{enumerate}[(a)]
	\item Continuity of sample paths,
	$$ \lim_{h\to0^+}\frac{1}{h}\, \mathbb{P} \,\, \Big({\big|X_{t+h}-X_t\big|>\delta}\big| X_t=x\Big)=0.$$
	\item Existence of a drift coefficient, i.e. there exists a function $f:\mathbb{R}\times[a,b]\longrightarrow \mathbb{R}$ such that
	$$\lim_{h\to0^+}\frac{1}{h}\, \mathbb{E} \,\,\Big(X_{t+h}-X_t\big| X_t=x\Big)=f(x, t).$$
	\item Existence of a diffusion coefficient, i.e. there exists a function $g:\mathbb{R}\times[a,b]\longrightarrow \mathbb{R}$ such that
	$$\lim_{h\to0^+}\frac{1}{h}\, \mathbb{E} \,\,\Big((X_{t+h}-X_t\big)^2| X_t=x\Big)=g^2(x,t).$$
\end{enumerate}
\end{definition}
The functions $f(x,t)$ and $g(x,t)$ are called, respectively, the {\it drift}
and {\it diffusion coefficients} of the diffusion process $(X_t)_{a\leq t \leq b}$.

\begin{definition} Let $h(x, t):\mathbb{R}\times[a,b]\longrightarrow \mathbb{R}$ be a measurable function.
	\begin{itemize}
		\item (Lipschitz condition) It is said that $h(x, t)$ satisfies the Lipschitz condition, in the first argument, if there exists a constant $C>0$ such that
		$|h(x, t)-h(y, t)| \leqslant C|x-y| \,\,\text {for all } \, x, y \in \mathbb{R} \,\,\text {and}\,\, a \leqslant t \leqslant b.$
		
		\item (Linear growth condition). It is said that $h(x, t)$ satisfies the linear growth condition, in the first argument, if there exists a constant $K>0$ such that
		$|h(x, t)| \leqslant K(1+|x|) \,\, \text {for all } \, x \in \mathbb{R} \,\,\text {and }\,\, a \leqslant t \leqslant b.$
	\end{itemize}
\end{definition}
\begin{definition}[Integration of smooth functions with respect to a diffusion process]\label{definition_HKI} Assume that $f(x,t)$ and $g(x,t)$ are continuous on $\mathbb{R}\times[a,b]$ and satisfy the Lipschitz and linear growth conditions in $x$. Consider a diffusion process $(X_t)_{t\geq 0}$ with drift $f(x,t)$ and diffusion coefficient $g(x, t)$. For any $\Phi\in \mathcal{C}^1(\mathbb{R},\mathbb{R})$, the H\"anggi-Klimontovich integral of $(\Phi(X_t))_{t\geq 0}$ in the interval $[a,b]$ with respect to the diffusion process $(X_t)_{t\geq 0}$ is defined as
	\begin{equation}\label{def_HKI}
	\int\limits_a^b \Phi(X_t)\bullet \mathrm{d}X_t :=\lim_{||\Delta_n||\to 0}\sum_{j=1}^{n}\Phi(X_{t_j})(X_{t_j}-X_{t_{j-1}}) \quad \text{in probability},
	\end{equation}
where $\Delta_n=\{t_0, t_1, \ldots, t_{n-1}, t_n\}$ is a partition of $[a, b]$, and $||\Delta_n||=\max_{1\leq j \leq n}(t_j-t_{j-1})$.
\end{definition}

Analogously to the previous case, in the following we will show that this integral is well-defined and verifies an identity that connects it to
the It\^o integral.

\begin{theorem}\label{thdefhki}
For any $\Phi\in \mathcal{C}^1(\mathbb{R},\mathbb{R})$, the limit in~\eqref{def_HKI} exists and moreover we have that the identity
\begin{equation}\label{hkid}
\int\limits_a^b \Phi(X_t)\bullet \mathrm{d}X_t=\int\limits_a^b \Phi(X_t)\,\mathrm{d}X_t+\int\limits_{a}^{b}\Phi^{\prime}(X_t)g^2(X_t,t)\,\mathrm{d}t\quad \text{holds almost surely.}
\end{equation}
\end{theorem}

\begin{proof}
The right-hand side of~\eqref{hkid} consists of the sum of an It\^o and a Lebesgue integral.
The It\^o integral is the limit
	\begin{equation}\label{def_II}
    \int\limits_a^b \Phi(X_t)\,\mathrm{d}X_t =\lim_{||\Delta_n||\to 0}\sum_{j=1}^{n}\Phi(X_{t_{j-1}})(X_{t_j}-X_{t_{j-1}}) \quad \text{in probability.}
    \end{equation}

Under the assumptions stated in this theorem, the It\^o integral is well-defined, see Chapter 5 in \cite{JeanFran}. On the other hand, the existence of the Lebesgue integral is also clear by the almost sure continuity of its integrand. As in the previous case, the existence of the limit in~\eqref{def_HKI} and the equality in~\eqref{hkid} can be proven altogether. In order to do so, we select the same $\Delta_n$-partitioning and consider the difference between the limit expressions on the right-hand sides
of Equation ~\eqref{def_HKI} and Equation ~\eqref{def_II}, denoted by $D_{\Delta_n}$.
Making use of the continuous differentiability of the function $\Phi(x)$ along with the Mean Value Theorem, we get
\begin{eqnarray}\nonumber
D_{\Delta_n} &=& \sum_{j=1}^{n}\Big[\Phi(X_{t_j})-\Phi(X_{t_{j-1}})\Big](X_{t_j}-X_{t_{j-1}}) \\ \nonumber
&=&\sum_{j=1}^{n}\Phi^{\prime}(\zeta_j)(X_{t_j}-X_{t_{j-1}})^2,
\end{eqnarray}
where $\zeta_j=(1-\theta_j)X_{t_{j-1}}+\theta_j X_{t_j}$ for fixed parameters $0 < \theta_j < 1$, $j=1,\cdots,n$. To complete the proof we have to check that
\begin{equation}\label{convergence_key}
\lim_{||\Delta_n||\to 0}\sum_{j=1}^{n}\Phi^{\prime}(\zeta_j)(X_{t_j}-X_{t_{j-1}})^2=\int\limits_{a}^{b}\Phi^{\prime}(X_t)g^2(X_t,t)\, \mathrm{d}t
\end{equation}
in probability. For this purpose, we bound
\begin{equation}\label{upper}
\sup_{1\leq j \leq n}\big| \Phi^{\prime}(\zeta_j) -\Phi^{\prime}(X_{t_{j-1}})\big|\leq \sup_{1\leq j \leq n}\Big(\sup_{x \in [X_{t_{j-1}}\wedge X_{t_j}, X_{t_{j-1}}\vee X_{t_j}]}\big| \Phi^{\prime}(x)-\Phi^{\prime}(X_{t_{j-1}})\big|\Big).
\end{equation}
As $(X_t)_{a\leq t \leq b}$ has continuous sample paths almost surely, $C:=\{X_t: t \in [a,b]\}$ is compact and $\Phi^{\prime}|_{C}$ is uniformly continuous. Thus, the right-hand side of Equation (\ref{upper}) tends to $0$ as $n \to \infty$ almost surely, and consequently $\varepsilon(n):=\sup_{1\leq j \leq n}\big| \Phi^{\prime}(\zeta_j) -\Phi^{\prime}(X_{t_{j-1}})\big|$ also tends to $0$ as $n \to \infty$ almost surely.
Furthermore, $\sum_{j=1}^{n}(X_{t_j}-X_{t_{j-1}})^2$ converges in probability to $\langle X\rangle_{t}|_{t=b}$, where $\langle X\rangle_{t}$ is the quadratic variation of $(X_t)_{a\leq t \leq b}$ (Proposition 4.21 in \cite{JeanFran}); this implies:
\begin{equation}\label{desig_diff}
\Bigg|\sum_{j=1}^{n}\Phi^{\prime}(\zeta_j)(X_{t_j}-X_{t_{j-1}})^2-\sum_{j=1}^{n}\Phi^{\prime}(X_{t_{j-1}})(X_{t_j}-X_{t_{j-1}})^2 \Bigg|\leq\varepsilon(n)\,\sum_{j=1}^{n}(X_{t_j}-X_{t_{j-1}})^2 \underset{n\to \infty}{\longrightarrow} 0
\end{equation}
in probability. As an intermediate step to prove the convergence (\ref{convergence_key}), it is necessary to verify that
\begin{equation}\label{key_diff}
\lim_{||\Delta_n||\to 0}\sum_{j=1}^{n}\Phi^{\prime}(X_{t_{j-1}})(X_{t_j}-X_{t_{j-1}})^2 =\int\limits_{a}^{b}\Phi^{\prime}(X_t)\,\mathrm{d}\langle X\rangle_{t} \quad \text{in probability}.
\end{equation}
In order to prove \eqref{key_diff}, we define a random measure $\mu_n$ on $[a,b]$ as follows:
$$
\mu_{n}(\mathrm{d} s):=\sum_{j=1}^{n}\left(X_{t_{j}}-X_{t_{j-1}}\right)^{2} \delta_{t_{j-1}}(\mathrm{d}s),
$$
where $\delta_{t}$ is the Dirac measure centered at $t$. Therefore, the sum on the left hand side of Equation (\ref{key_diff}) can be expressed as
$$
\sum_{j=1}^{n} \Phi^{\prime}\left(X_{t_{j-1}}\right)\left(X_{t_{j}}-X_{t_{j-1}}\right)^{2}=\int_{[a,b]} \Phi^{\prime}\left(X_{s}\right) \mu_{n}(\mathrm{d} s).
$$
As a consequence of Proposition 4.21 in \cite{JeanFran}, we get for every $s \in \Delta_n$ and any $n \ge 1$
\begin{equation*}
 \mu_{n}([a, s]) \underset{n \rightarrow \infty}{\longrightarrow}\langle X\rangle_{s} \quad \text {in probability},
\end{equation*}
which implies that the sequence of measures $\mu_{n}$ converges vaguely (see Chapter 4 in \cite{kallenberg}) to the measure
$\mathlarger{\mathlarger{\mathbbm{1}}}_{[a, b]}(s)\,\mathrm{d}\langle X\rangle_{s}$ in probability. Thus, since $\Phi \in \mathcal{C}^{1}(\mathbb{R}, \mathbb{R})$,
we use Lemmata 4.1 and 4.8, and Corollary 4.9 from \cite{kallenberg} to find that
\begin{equation}\label{convergence_rm}
 \int_{[a, b]} \Phi^{\prime}\left(X_{s}\right) \mu_{n}(\mathrm{d} s) \underset{n \rightarrow \infty}{\longrightarrow} \int_{a}^{b} \Phi^{\prime}\left(X_{s}\right) \mathrm{d}\langle X\rangle_{s}\quad \text {in probability}.
\end{equation}

Since the convergence in \eqref{convergence_rm} implies \eqref{key_diff}, to complete the proof we just need to compute $\langle X\rangle_{t}$. To this end, according to Theorem 10.8.9 in \cite{kuo}, the diffusion process $(X_t)_{a\leq t \leq b}$ solves the stochastic integral equation
$$
X_{t}=X_{a}+\int_{a}^{t} f\left(X_{s}, s\right) \mathrm{d} s+\int_{a}^{t} g\left(X_{s}, s\right) \mathrm{d} W_{s}, \quad a \leqslant t \leqslant b.
$$
So, the quadratic variation of $(X_t)_{a\leq t \leq b}$ is
$$
\begin{aligned}
	\langle X\rangle_{t} =\int_{a}^{t} g^{2}\left(X_{s}, s\right) \mathrm{d}\langle W\rangle_{s}=\int_{a}^{t} g^{2}\left(X_{s}, s\right) \mathrm{d}s,
\end{aligned}
$$
or, employing the stochastic differential notation, $\mathrm{d} \langle X\rangle_{t}=g^2(X_t,t)\,\mathrm{d}t$.
Substituting this result in \eqref{key_diff} yields (\ref{convergence_key}), and therefore the existence of the H\"anggi-Klimontovich integral.
Finally, by taking an appropriate subsequence if necessary, the identity in Equation \eqref{hkid} is verified.
\end{proof}

\begin{remark}
As noted in the previous section, we may summarize the equality in the statement by employing the stochastic differential notation $\Phi(X_t)\bullet \mathrm{d}X_t=\Phi(X_t)\,\mathrm{d}X_t+\Phi^{\prime}(X_t)g^2(X_t,t)\,\mathrm{d}t$.
\end{remark}

\section{Multidimensional Generalization}\label{multidhki}

To develop a theory of stochastic differential equations based on the H\"anggi-Klimontovich interpretation, we still need to further extend our results from section \ref{One_HKI}. To this end, we now consider generalizations of this integral for multidimensional diffusion processes. Following Stratonovich \cite{stratonovich}, no theory for stochastic differential equations can be built from the results of the previous section, and the full multidimensional integral is needed to address such models.

\begin{definition} 	Let $m$ and $d$ be positive integers and $\mathcal{M}_{d\times m}(\mathbb{R})$ is the set of all $d\times m$ matrices with real coefficients. Let $\rho:\mathbb{R}^m\times[a,b] \longrightarrow \mathbb{R}^m$ and $B:\mathbb{R}^m\times[a,b] \longrightarrow \mathcal{M}_{d\times m}(\mathbb{R})$ be two measurable functions.

	\begin{itemize}
		\item (Lipschitz condition) It is said that $\rho (x, t)$ and $B(x, t)$ satisfy the Lipschitz condition, in the first argument, if there exists a constant $C>0$ such that for all $x, y \in \mathbb{R}^m$ and $a \leqslant t \leqslant b,$
		$$|\rho(x,t)-\rho(y,t)|\leqslant C|x-y|,\qquad ||B(x, t)-B(y, t)||_{HS} \leqslant C|x-y|.$$
		\item (Linear growth condition). It is said that $\rho (x, t)$ and $B(x, t)$ satisfy the linear growth condition, in the first argument, if there exists a constant $K>0$ such that for all $x \in \mathbb{R}^m$ and $a \leqslant t \leqslant b,$
		$$|\rho(x,t)| \leqslant K(1+|x|),\qquad ||B(x, t)||_{HS} \leqslant K(1+|x|).$$
	\end{itemize}
	Here $|x|$ denotes the Euclidean norm of $x \in \mathbb{R}^m$. Furthermore, the Hilbert-Schmidt norm of a $d\times m$ matrix $B=[b_{ij}]$ is defined as
	$$||B||_{HS}:=\bigg(\sum_{i=1}^{d}\sum_{j=1}^{m} b^2_{ij}\bigg)^{1/2}.$$
\end{definition}

A Markov process $\mathbf{X}_t = (X^{(1)}_t, X^{(2)}_t,\ldots, X^{(m)}_t)$ is called an $\mathbb{R}^m$-valued diffusion process,
described by the drift vector $\rho(x,t)$ and the diffusion coefficient matrix $B(x,t)$, if it componentwise satisfies three conditions analogous to those established in Definition \ref{definition_diffusion}. For an explicit formulation see Definition 10.8.3 in \cite{kuo} or Definition 5.2.1 in \cite{Pavliotis}.

\begin{definition}\label{def_HKI_multi}
	Assume that $\rho(x,t)$ and $B(x,t)=[b_{ij}(x,t)]$ are continuous on $\mathbb{R}^m\times[a,b]$ and both satisfy the Lipschitz and linear growth conditions in $x$. Also, consider a $m$-dimensional diffusion process $(\mathbf{X}_t)_{t\geq 0}$, with drift vector $\rho(x,t)$ and diffusion coefficient matrix $B(x,t)$. For any continuous $\Psi:\mathbb{R}^m\times[a,b]\longrightarrow\mathcal{M}_{d\times m}(\mathbb{R})$, for which all its partial derivatives
$\partial_{x_i}\Psi(x,t)$ exist and are continuous, the H\"anggi-Klimontovich integral of $\big(\Psi(\mathbf{X}_t,t)\big)_{t \geq 0}$ in the interval $[a,b]$ with respect to the diffusion process $(\mathbf{X}_t)_{\geq 0}$ is defined as
		\begin{equation}\label{HKI_multi}
		\int\limits_a^b \Psi(\mathbf{X}_t,t)\bullet \mathrm{d}\mathbf{X}_t :=\lim_{||\Delta_n||\to 0}\sum_{j=1}^{n}\Psi(\mathbf{X}_{t_j},t_j)(\mathbf{X}_{t_j}-\mathbf{X}_{t_{j-1}}) \quad \text{in probability},
	\end{equation}
	where $\Delta_n=\{t_0, t_1, \ldots, t_{n-1}, t_n\}$ is a partition of $[a, b]$, and $||\Delta_n||=\max_{1\leq j \leq n}(t_j-t_{j-1})$.
\end{definition}

\begin{remark}
The products in this definition are clearly scalar, so that the integral yields an $\mathbb{R}^d$-valued random variable.
\end{remark}

This integral is well-defined and its relationship with the Itô integral is expressed in the following theorem.

\begin{theorem}\label{Theorem_RCM}
	Under the assumptions established in Definition \ref{def_HKI_multi}, the limit in Equation \eqref{HKI_multi} exists. Furthermore, it is connected to the Itô integral through the following identity
	\begin{equation}\label{RCM}
		\int\limits_a^b \Psi(\mathbf{X}_t,t)\bullet \mathrm{d}\mathbf{X}_t=\int\limits_a^b \Psi(\mathbf{X}_t,t)\,\mathrm{d}\mathbf{X}_t+\sum_{l=1}^{m}\sum_{k=1}^{m}\int\limits_a^b\big(\partial_{x_k}\Psi(\mathbf{X}_t,t)\big)_{* l} b_{lk}(\mathbf{X}_t,t)\,\mathrm{d}t,
	\end{equation}
which holds almost surely. Here, the $d$-dimensional vector $\big(\partial_{x_k}\Psi\big)_{* l}$ denotes the $l$'th column of the $d\times m$ matrix $\partial_{x_k}\Psi=\Big(\frac{\partial\Psi_{ij}}{\partial_{x_k}}\Big)$.
\end{theorem}

\begin{proof}
	The proof is analogous to that of the one-dimensional case. The first summand on the right-hand side of Equation \eqref{RCM} corresponds to the Itô
integral that in this case is defined as the limit
		\begin{equation}\label{II_multi}
		\int\limits_a^b \Psi(\mathbf{X}_t,t)\, \mathrm{d}\mathbf{X}_t =\lim_{||\Delta_n||\to 0}\sum_{j=1}^{n}\Psi(\mathbf{X}_{t_{j-1}},t_j)(\mathbf{X}_{t_j}-\mathbf{X}_{t_{j-1}}) \quad \text{in probability}.
	\end{equation}
Its existence is again guaranteed by classical stochastic analysis theory. On the other hand, the second summand corresponds to a sum of Lebesgue integrals that are all well-defined by the continuity of their integrands. Now select the same $\Delta_n$-partitioning and consider the difference between the summations in Equation \eqref{HKI_multi} and Equation \eqref{II_multi} to get
\begin{eqnarray}\nonumber
	D_{\Delta_n} = \sum_{j=1}^{n}\sum_{l=1}^{m}\Big[\Psi_{* l}(\mathbf{X}_{t_j},t_j)-\Psi_{* l}(\mathbf{X}_{t_{j-1}},t_j)\Big]\big({X}_{t_j}^{(l)}-{X}_{t_{j-1}}^{(l)}\big).
\end{eqnarray}
Making use of the continuous differentiability of the function $\Psi(x,t)$ along with the Mean Value Theorem we have that
\begin{eqnarray}\nonumber
	D_{\Delta_n} &=& \sum_{j=1}^{n}\sum_{l=1}^{m}\Big[\big(\nabla \Psi_{* l}(\xi_j,t_j)\big)^\top\big(\mathbf{X}_{t_j}-\mathbf{X}_{t_{j-1}}\big)\Big]\big({X}_{t_j}^{(l)}-{X}_{t_{j-1}}^{(l)}\big) \\ \nonumber
	&=&\sum_{j=1}^{n}\sum_{l=1}^{m}\sum_{k=1}^{m}\Bigg(\frac{\partial\Psi(\xi_j,t_j)}{\partial{x_k}}\Bigg)_{* l}\big({X}_{t_j}^{(k)}-{X}_{t_{j-1}}^{(k)}\big)\big({X}_{t_j}^{(l)}-{X}_{t_{j-1}}^{(l)}\big),
\end{eqnarray}
where $\nabla \Psi$ represents the gradient of $\Psi$ and $\xi_j \in [\mathbf{X}_{t_{j-1}}, \mathbf{X}_{t_j}]$ for all $j=1, 2,\cdots,n$. Using a slight modification of the arguments in the one-dimensional case, for $l$ and $k=1,2, \cdots,m$, it is verified that the following convergence in probability
\begin{equation}\label{convergence_mc}
\sum_{j=1}^{n} \Bigg(\frac{\partial\Psi(\xi_j,t_j)}{\partial{x_k}}\Bigg)_{* l}\big({X}_{t_j}^{(k)}-{X}_{t_{j-1}}^{(k)}\big)\big({X}_{t_j}^{(l)}-{X}_{t_{j-1}}^{(l)}\big) \underset{n \rightarrow \infty}{\longrightarrow} \int\limits_a^b\big(\partial_{x_k}\Psi(\mathbf{X}_t,t)\big)_{* l}\,\mathrm{d}\langle X^{(l)}, X^{(k)}\rangle_{t}
\end{equation}
holds, where $\langle X^{(l)}, X^{(k)}\rangle_{t}$ is the cross-variation process of $X^{(l)}_t$ and $X^{(k)}_t$. We calculate $\langle X^{(l)}, X^{(k)}\rangle_{t}$ analogously to the one-dimensional case to find
\begin{equation}\label{cross_variation}
	\mathrm{d}\langle X^{(l)}, X^{(k)}\rangle_{t}=b_{lk}(\mathbf{X}_t,t)\,\mathrm{d}t.
\end{equation}
Thus, from \eqref{convergence_mc} and \eqref{cross_variation}, we conclude that
$$\lim\limits_{||\Delta_n||\to 0}D_{\Delta_n} =\sum_{l=1}^{m}\sum_{k=1}^{m}\int\limits_a^b\big(\partial_{x_k}\Psi(\mathbf{X}_t,t)\big)_{* l}\,b_{lk}(\mathbf{X}_t,t)\,\mathrm{d}t \quad \text{in probability}.$$
Therefore, from this result, the existence of the limit in Equation \eqref{HKI_multi} is deduced and, by taking an appropriate subsequence, the identity in Equation \eqref{RCM} is verified.
\end{proof}

\begin{remark}
This result is key to building a theory for HK stochastic differential equations, as the following section will show. But it is also fundamental to understand the connection of the Fick law of diffusion with the stochastic trajectories that underlie it. Such a connection is both technical and physical, and has been studied in \cite{escudero2025beneath}.
\end{remark}

\section{H\"anggi-Klimontovich Stochastic Differential Equations}\label{sechkisdes}

\begin{definition}\label{defsolhksde}
We say that a diffusion process $X_t$ that satisfies a stochastic integral equation of the form
\begin{equation}
X_t=x_0+\int\limits_{0}^t f(X_s, s)\,\mathrm{d}s+\int\limits_{0}^t g(X_s, s)\bullet\mathrm{d}W_s,
\end{equation}
for $0 \le t \le T$, for some real numbers $x_0$ and $T$, and for some functions $f$ and $g$ such that all
the expressions are well-defined,
is a solution to the H\"anggi-Klimontovich stochastic differential equation (HK SDE)
\begin{equation}\label{HK-SDE}
\mathrm{d}X_t= f(X_t, t)\,\mathrm{d}t+g(X_t, t)\bullet\mathrm{d}W_t, \qquad X_0=x_0,
\end{equation}
in the interval $t \in [0,T]$.
\end{definition}

The following statement identifies conditions on the functions $f$ and $g$ that guarantee that all the expressions in Definition \ref{defsolhksde}
are well-defined, and moreover that a unique solution to the HK SDE exists.

\begin{proposition}[Conversion rule for HK SDEs]\label{proptrans}
Let $f:\mathbb{R}\times[0, T] \longrightarrow \mathbb{R}$ and  $g:\mathbb{R}\times[0, T] \longrightarrow \mathbb{R}$ be functions that fulfill the properties specified in Definition \ref{definition_HKI}. Moreover assume that $g$ is continuously differentiable and that $\frac{\partial g}{\partial x}g$ satisfies the Lipschitz and linear growth conditions in $x$. Then the unique solution to the It\^o SDE
\begin{equation}\label{transformation_SDE}
\mathrm{d}X_t= \Big[f(X_t, t)+\frac{\partial g}{\partial x}(X_t,t)g(X_t, t)\Big]\,\mathrm{d}t+ g(X_t, t)\,\mathrm{d}W_t, \qquad X_0=x_0,
\end{equation}
solves Equation~\eqref{HK-SDE} almost surely. Correspondingly, there almost surely exists a unique solution to~\eqref{HK-SDE} that is given
by the solution to~\eqref{transformation_SDE}.
\end{proposition}

\begin{proof}
Under the stated assumptions, the SDE~\eqref{transformation_SDE} has a unique solution~\cite{kuo,oksendal}; our first step is to prove that it
solves Equation~\eqref{HK-SDE}. To this end rewrite the stochastic integral in~\eqref{HK-SDE} as
\begin{equation*}
\int\limits_{0}^{t} g(X_s,s)\bullet \mathrm{d}W_s =	\int\limits_{0}^{t}\big(0, g(X_s,s)\big)\bullet \mathrm{d} \begin{pmatrix} X_s  \\ W_s \end{pmatrix}.
\end{equation*}
If we take $X_t$ to be the solution to SDE~\eqref{transformation_SDE}, we recognize in the right-hand side of this equation a particular case of
the stochastic integral introduced in Definition~\ref{def_HKI_multi}; to be concrete we have selected $d=1$, $m=2$, $X_t^{(1)}=X_t$, $X_t^{(2)}=W_t$,
and $\Psi=\big(0,g\big)$. Classical stochastic analytical results~\cite{kuo,oksendal} guarantee that $(X_t,W_t)$ is a diffusion process with
the desired properties, so we fall under the hypotheses of Theorem \ref{Theorem_RCM}, and therefore may use transformation formula~\eqref{RCM} to find
\begin{equation*}
\int\limits_{0}^{t}\big(0, g(X_s,s)\big)\bullet \mathrm{d} \begin{pmatrix} X_s  \\ W_s \end{pmatrix}
= \int\limits_{0}^t g(X_s,s)\, \mathrm{d}W_s + \int\limits_{0}^t \frac{\partial g}{\partial x}(X_s,s)g(X_s, s)\,\mathrm{d}s.
\end{equation*}
This automatically implies that the unique solution of~\eqref{transformation_SDE} solves~\eqref{HK-SDE} almost surely. To check uniqueness assume there exist two solutions; by the same formula this enters in contradiction with the uniqueness of solution to Equation~\eqref{transformation_SDE}.
\end{proof}

A direct consequence of this proposition is the following

\begin{corollary}
Let $f:\mathbb{R}\times[0, T] \longrightarrow \mathbb{R}$ and  $g:\mathbb{R}\times[0, T] \longrightarrow \mathbb{R}$ be functions that fulfill the properties specified in Definition \ref{definition_HKI}. Moreover assume that $g$ is continuously differentiable and that $\frac{\partial g}{\partial x}g$ satisfies the Lipschitz and linear growth conditions in $x$. Then the unique solution to the It\^o SDE
\begin{equation}\label{transformation_SDE2}
\mathrm{d}X_t= f(X_t, t)\,\mathrm{d}t+ g(X_t, t)\,\mathrm{d}W_t, \qquad X_0=x_0,
\end{equation}
solves the equation
\begin{equation}\nonumber
\mathrm{d}X_t= \Big[f(X_t, t)-\frac{\partial g}{\partial x}(X_t,t)g(X_t, t)\Big] \, \mathrm{d}t+ g(X_t, t) \bullet \mathrm{d}W_t, \qquad X_0=x_0,
\end{equation}
almost surely. Correspondingly, there almost surely exists a unique solution to this equation that is given
by the solution to~\eqref{transformation_SDE2}.
\end{corollary}

\begin{remark}
From Proposition ~\ref{proptrans} and its corollary it follows that, if $g$ is independent of $X_t$, i.e. $g(x,t)=g(t)$,
Equation~\eqref{transformation_SDE} and Equation~\eqref{transformation_SDE2} reduce to the same SDE.
Moreover, since under the present assumptions the existence and uniqueness of solution is granted,
this means that the two interpretations of the stochastic differential equation, Itô and H\"anggi-Klimontovich, lead to the same solution.
This is not surprising since, under this hypothesis, both definitions of stochastic integral coincide.
\end{remark}

\section{Fokker-Planck Equation and Asymptotic Behavior}\label{fpeab}

It is central to the theory of diffusion processes that their distribution function obeys a second order linear parabolic partial differential
equation: the Fokker-Planck equation. This equation has been the main object of study in an important part of the physical literature, rather
than SDEs or the paths of the diffusion process. Indeed, part of the importance of the H\"anggi-Klimontovich interpretation relies on the
properties of the Fokker-Planck equation it generates. The following proposition assures the existence of a suitable Fokker-Planck equation
associated to HK SDEs.

\begin{proposition}\label{FPE_P}
    Let $X_t$ be a diffusion process that satisfies the HK SDE~\eqref{HK-SDE}. Assume that $f(x,t)$ and $g(x,t)$ obey the same assumptions as in Proposition~\ref{proptrans} and moreover that the partial derivatives $\frac{\partial f}{\partial x}$, $\frac{\partial g}{\partial x}$, and $\frac{\partial^2g}{\partial x^2}$ satisfy the Lipschitz and linear growth conditions in $x$. Assume in addition that there exists a constant $c>0$ such that $g(x,t)\geq c  \,\, \text {for all } \, x \in \mathbb{R} \,\,\text {and }\,\, 0 \leqslant t \leqslant T.$ Then, the transition probability density $p(x, t|x_0, t_0)$, with $0 \leqslant t_0 < t$ and $x_0 \in \mathbb{R}$, of the process $X_t$ is given by the unique solution of the Fokker-Planck Equation (FPE)
    \begin{equation}\label{FPE1}
\partial_{t} p(x, t|x_0,t_0)=-\partial_{x}\big[f(x,t)+ \partial_{x}g(x,t)g(x,t)\big]p(x,t|x_0,t_0)+\frac{1}{2}\partial_{xx}g^2(x, t)p(x,t|x_0,t_0),
\end{equation}
with initial condition $$\lim\limits_{t\to t_0+}p(x,t|x_0,t_0)=\delta(x_0-x),$$
where $\delta(\cdot)$ denotes the Dirac measure.
\end{proposition}
\begin{proof}
 The statement is a consequence of Proposition \ref{proptrans} and Theorem 10.9.10 in~\cite{kuo}. First, Proposition \ref{proptrans} guarantees the
 existence of a unique diffusion process that solves the HK SDE~\eqref{HK-SDE}. Furthermore, this proposition establishes this process
 as the unique solution to a particular It\^o SDE; the application of Theorem 10.9.10 from~\cite{kuo} to this SDE completes the proof.
\end{proof}

From now on we assume that $X_t$ is a homogeneous diffusion process, that is, its drift and diffusion coefficients are independent of time;
in symbols: $f(x,t)=f(x)$ and $g(x,t)=g(x)$. Under these assumptions, the FPE \eqref{FPE1} reduces to
\begin{equation}\label{continuity_equation}
	\partial_{t}p_t(x_0,x)+\partial_{x}J(x)=0 \quad\textrm{with} \quad \lim\limits_{t\to 0+}p_t(x_0,x)=\delta(x_0-x),
\end{equation}
where
\begin{equation*}
	J(x):= \big[f(x)+ \partial_{x}g(x)g(x)\big]p_t(x_0,x)-\frac{1}{2}\partial_{x}g^2(x)p_t(x_0,x).
\end{equation*}
Equation \eqref{continuity_equation} has the form of a local conservation law in which $J(x)$ can be interpreted as a probability flow; therefore it
has to be supplemented with suitable boundary conditions \cite{Gardiner,HL84,Pavliotis}. Next, we calculate the stationary solution of the homogeneous Fokker-Planck equation with reflecting boundary conditions on a finite domain, since these conditions combine their intuitive meaning,
physical applicability, and mathematical tractability. Other types of boundary conditions can be seen in \cite{Gardiner,HL84,Pavliotis}.

\begin{proposition}\label{prop_stationary}
The homogeneous diffusion process $X_t$ admits a stationary distribution $p_s(x)$ when constrained to the finite interval $[a,b]$ with reflecting borders.
Moreover, $p_s(x)$ is the unique time-independent probability distribution that solves the Fokker-Planck equation \eqref{continuity_equation} on $[a, b]$
and satisfies the boundary conditions $J(a)=J(b)=0$; furthermore, it is given by the explicit formula
    \begin{equation}\label{Stationary_Function}
		p_s(x)=\mathcal{N}_0\,e^{\mathcal{V}(x)},
	\end{equation}
where $\mathcal{V}(x):=2\int\limits_{a}^{x} f(u)/g^2(u) \,\mathrm{d}u$ and\,
$\mathcal{N}_0:= \left( \int\limits_{a}^{b}\exp[\mathcal{V}(u)]\,\mathrm{d}u \right)^{-1}$ is the normalization constant.
\end{proposition}

\begin{proof}
First, we compute the explicit solution formula. Since for any time-independent solution $\partial_{t}p_s(x)=0$, then Equation \eqref{continuity_equation} can be written in terms of the probability flow as $\partial_{x}J(x)=0$. Therefore $J(x)$ is constant and, moreover, the reflecting boundary conditions
imply $J(x)=0$ for $x \in [a,b]$. So the problem reduces to the ordinary differential equation
\begin{equation*}\label{edo1}
	\big[f(x)+ g^\prime(x)g(x)\big]p_s(x)=\frac{1}{2}\,\frac{\mathrm{d}}{\mathrm{d}x}\big[g^2(x)p_s(x)\big],
\end{equation*}
which is readily solvable to yield
\begin{equation*}\label{StationaryD}
	p_s(x)=\mathcal{N}_0\,\textrm{exp}\,\Bigg\{2\int\limits_{a}^{x}\frac{f(u)}{g^2(u)}\,\mathrm{d}u\Bigg\}.
\end{equation*}
The classical theory of ordinary differential equations guarantees that this one-parameter family of solutions, indexed by $\mathcal{N}_0$,
is the unique solution set to this equation. If this solution has to be a probability distribution, then it has to be normalizable.
Under the present assumptions, the parameter can be assimilated to a normalization constant, which is clearly finite and given by
$\mathcal{N}_0=\Big(\int\limits_{a}^{b}\exp[\mathcal{V}(u)]\,\mathrm{d}u\Big)^{-1}<\infty$; this fixes the uniqueness of solution.
\end{proof}

\begin{remark}
The real importance of the existence and uniqueness of $p_s(x)$ is the fact that it is a global attractor of the time-dependent solutions $p_t(x_0, x)$ of \eqref{continuity_equation}. Such a global stability is guaranteed by the existence of a Lyapunov functional. To be precise, every solution $p_t(x_0, x)$ to the Fokker-Planck equation \eqref{continuity_equation}, with arbitrary initial condition, converges in relative entropy
(or Kullback–Leibler divergence) $H(t)$ to $p_s(x)$ in the long-time limit, i.e.
$$H(t):=\int\limits_{a}^{b}p_t(x_0, x)\ln\bigg(\frac{p_t(x_0, x)}{p_s(x)}\bigg)\,\mathrm{d}x \underset{t \rightarrow \infty}{\longrightarrow}0.$$
This result is also a consequence of Proposition \ref{proptrans} and Theorem 10.9.10 in~\cite{kuo} in very much the same way as the proof of
Proposition \ref{FPE_P}, along with the developments in Chapter 6 of \cite{HL84}.
\end{remark}

\begin{remark}\label{rem:noneqpot}
The function $\mathcal{V}(x)=2\int\limits_{a}^{x} f(u)/g^2(u) \,\mathrm{d}u$ is sometimes referred to as nonequilibrium potential.
Using this terminology we find the appealing form for the stationary probability distribution $p_s(x) \propto \exp[\mathcal{V}(x)]$.
This structure has been viewed as one of the strong points of the HK interpretation of noise in a fraction of the physical literature. However,
the same structure results in the Itô and Stratonovich cases upon a redefinition of the nonequilibrium potential, see \cite{HL84}. So the only real
advantage is the comparatively simpler appearance of the nonequilibrium potential in the HK case; notwithstanding, this simplicity is sometimes
regarded as a feature of its superiority.
\end{remark}

The following result is considered as another advantage of the HK interpretation in part of the physical literature.

\begin{theorem}[Robustness of deterministic behavior]\label{linear_stability}
	The set of critical points of the stationary distribution \eqref{Stationary_Function}, i.e. $\{x \in (a,b): p^{\prime}_s(x)=0\}$,
coincides with the set of fixed points of the dynamical system generated by the ordinary differential equation
	\begin{equation}\label{edo}
	\frac{\mathrm{d}x_t}{\mathrm{d} t}=f(x_t).
\end{equation}
Furthermore, the non-degenerate local maxima of \eqref{Stationary_Function} coincide with the stable fixed points of \eqref{edo}. Correspondingly,
the non-degenerate local minima of \eqref{Stationary_Function} coincide with the unstable fixed points of \eqref{edo}.
\end{theorem}

\begin{remark}
Note that Equation \eqref{edo} is nothing but the deterministic counterpart of HK SDE \eqref{HK-SDE} under the time homogeneity
assumption and with $g \equiv 0$.
\end{remark}

\begin{proof}
First of all note that under the current assumptions, equation \eqref{edo} presents unique and global-in-time solutions to its associated initial value
problems; therefore it generates a well-defined dynamical system.
Again these assumptions guarantee that both $p_s(x)$ and $\mathcal{V}(x)$ are differentiable; thus by direct differentiation of \eqref{Stationary_Function} we establish
$$\{x: p^{\prime}_s(x)=0\} = \{x:\mathcal{V}^{\prime}(x)=0\},$$
and also by direct differentiation, this time of the definition of $\mathcal{V}(x)$, it follows that
$$\{x:\mathcal{V}^{\prime}(x)=0\}=\{x:f(x)=0\}.$$
Both equalities united yield the first claim in the statement.

Now define $\Gamma:=\{x:f(x)=0\}$ and differentiate \eqref{Stationary_Function} once more to find
	\begin{equation}\label{curvature_equation}
	p_s^{\prime\prime}(x)=\frac{f^{\prime}(x)}{g^2(x)}\,\mathcal{N}_0\,e^{\mathcal{V}(x)} \quad \textrm{for all}\quad x\in \Gamma.
	\end{equation}
	From \eqref{curvature_equation} we conclude that $\{x\in \Gamma :p_s^{\prime\prime}(x)<0\}=\{x\in\Gamma:f^{\prime}(x) <0\}$, i.e. the set
of local maxima of $p_s(x)$ coincides with the set of stable fixed points of \eqref{edo}. Similarly, we have that
$\{x\in\Gamma:p_s^{\prime\prime}(x)>0\}=\{x\in\Gamma:f^{\prime}(x) >0\}$, so the last claim in the statement follows.
\end{proof}

\begin{remark}\label{rem:linest}
Theorem \ref{linear_stability} establishes a connection between the steady-state behavior of a deterministic dynamical system governed by equation \eqref{edo} and the stationary distribution of its stochastic counterpart, provided it is HK-interpreted. Precisely, the local maxima of the
stationary distribution, which denote the states of the physical system that are more frequently observed locally (at least in the long time),
coincide identically with the stable fixed points of the dynamical system; respectively, the local minima of this distribution,
which correspond to the physical states that are more rarely observed locally in an experiment, equal the unstable fixed points of the dynamical system.
This sort of robustness of the deterministic behavior is the consequence of the H\"anggi-Klimontovich interpretation. Other interpretations, such as Itô or Stratonovich, do not possess this property, see Chapter 6 in \cite{HL84}. On the other hand, it is easy to see that additive noise does share this
property with multiplicative HK noise. Note that this does not mean in either case that the solution to the stochastic equation is simply
the noisy counterpart of the solution to the deterministic equation. For instance, transitions between different stable states, which are forbidden
in the absence of noise, may happen in the presence of stochastic forcing.
\end{remark}

We finish this section emphasizing, once more, that the definitive test for any nonequilibrium potential is the comparison to physical stationary
probability distributions. In this respect, a possible starting point to unveil possible advantages of the different stochastic calculi is to depart from recent developments in that field \cite{dhawan2024ito,lente2025direction}. In particular, a recent study on how to select the interpretation of noise based on experimental data is~\cite{pacheco2024langevin}.

\section{Applications and Difficulties}\label{secappdiff}

While one can frequently find in the physical literature references to the appealing properties of the H\"anggi-Klimontovich interpretation that
make it apparently better suited to applications in the field than other stochastic integrals, it is not free from problems either. In the following, we
discuss several physical systems for which the HK integral presents difficulties that are not present for the It\^o or even the Stratonovich
interpretations.

\subsection{Kinetic energy of one Langevin particle}

The velocity $V_t$ of a point Brownian particle with mass $m>0$, surrounded by a homogeneous heat bath of constant temperature,
obeys the Langevin equation \cite{langevin}
\begin{equation}\label{langevin1}
m\,\mathrm{d}V_t= - \gamma\, V_t\, \mathrm{d}t + \sigma\,\mathrm{d}W_t, \quad \quad V_0=v_0,
\end{equation}
where $\gamma>0$ denotes the friction constant of the surrounding medium and $\sigma >0$ is the amplitude of the thermal fluctuations;
moreover, we assume for simplicity that the initial velocity $v_0$ is a real number and that all parameters are fixed (for the limit of a vanishing mass, that is $m \downarrow 0$, see~\cite{cescudero}). We note that this model is also meaningful in finance \cite{vasicek}.
Obviously, the unique solution to this equation is an Ornstein–Uhlenbeck process \cite{uo}. The kinetic energy of this particle is
\begin{equation*}
K_t=\frac{1}{2}mV_t^2.
\end{equation*}
It turns out that this quantity has been used as a benchmark for the different stochastic calculi both classically \cite{Kampen}
and in more recent times \cite{cescudero}.
To obtain the stochastic differential equation that governs the evolution of $K_t$ we shall consider three approaches. Using Itô
lemma (see for instance \cite{Kampen}) we get that
\begin{equation}\label{kito}
\mathrm{d} K_t = \frac{\sigma^2}{2m} \mathrm{d}t -2\frac{\gamma}{m} \, K_t \, \mathrm{d}t + \sqrt{2 \frac{\sigma^2}{m} \, K_t} \, \mathrm{d}W_t,
\qquad K_0=\frac{1}{2}mv_0^2;
\end{equation}
while using Stratonovich calculus, as done for instance in \cite{Kampen}, we find that
\begin{equation}\label{strat}
\mathrm{d} K_t =  -2\frac{\gamma}{m} \, K_t \, \mathrm{d}t + \sqrt{2 \frac{\sigma^2}{m} \, K_t} \circ \mathrm{d}W_t,
\qquad K_0=\frac{1}{2}mv_0^2.
\end{equation}
Alternatively, to use the H\"anggi-Klimontovich calculus, we will employ the conversion rule in Theorem \ref{Theorem_RCM}. To this end, if $V_t$ is the solution to SDE \eqref{langevin1}, we can identify, in the right-hand side of this equation, a particular case of the stochastic integral introduced in Definition~\ref{def_HKI_multi}. In particular, select $d=1$, $m=2$, $X_t^{(1)}=V_t$, $X_t^{(2)}=W_t$, and $\Psi=\big(0,x\big)$ in that definition. The classical theory \cite{kuo,oksendal} guarantees that $(V_t,W_t)$ is a diffusion process with the necessary properties, so it falls under the hypotheses of Theorem \ref{Theorem_RCM}. Then use the transformation formula \eqref{RCM} to get
\begin{equation*}
\int\limits_{0}^{t} V_s \bullet \mathrm{d} W_s
=\int\limits_{0}^{t}\big(0, V_s\big)\bullet \mathrm{d} \begin{pmatrix} V_s  \\ W_s \end{pmatrix}
= \int\limits_{0}^t V_s \, \mathrm{d}W_s + \int\limits_{0}^t \frac{\sigma}{m}\,\mathrm{d}s,
\end{equation*}
or, in differential notation,
\begin{equation}\nonumber
V_t \, \bullet \mathrm{d}W_t =  V_t \, \mathrm{d}W_t + \frac{\sigma}{m} \, \mathrm{d}t.
\end{equation}
This relation, in conjunction with the definition of $K_t$ and \eqref{kito}, leads to the equation
\begin{equation}\label{khk}
\mathrm{d} K_t = -\,\frac{\sigma^2}{2m} \,\mathrm{d}t -2\frac{\gamma}{m} \, K_t \, \mathrm{d}t + \sqrt{2 \frac{\sigma^2}{m} \, K_t} \, \bullet \mathrm{d}W_t,
\qquad K_0=\frac{1}{2}mv_0^2.
\end{equation}
Note that the Stratonovich equation \eqref{strat} could have been derived analogously from the relation $V_t \, \circ \mathrm{d}W_t =  V_t \, \mathrm{d}W_t + [\sigma/(2m)] \, \mathrm{d}t$.
Among these three interpretations, the Itô equation \eqref{kito} possesses a unique solution which is both strong and global, as follows from the Watanabe--Yamada theorem \cite{wy}, see \cite{cescudero,cescudero2}. On the other hand, the Stratonovich equation \eqref{strat} has infinitely many spurious solutions \cite{cescudero,cescudero2}.
This follows from the fact that the state $K_t=0$ might be absorbing for this equation. This becomes clearer if we consider the particle to be initially
at rest, that is
\begin{equation}\label{strat2}
\mathrm{d} K_t =  -2\frac{\gamma}{m} \, K_t \, \mathrm{d}t + \sqrt{2 \frac{\sigma^2}{m} \, K_t} \circ \mathrm{d}W_t,
\qquad K_0=0.
\end{equation}
This equation admits the solution $K_t=0$ for all times, i.e. the particle remains at rest forever. Of course, that does not make any physical sense since
it would \emph{de facto} imply that fluctuations due to the thermal bath would have permanently disappeared. If we established the same initial condition for the H\"anggi-Klimontovich case
\begin{equation}\label{khk2}
\mathrm{d} K_t = -\,\frac{\sigma^2}{2m}\, \mathrm{d}t -2\frac{\gamma}{m} \, K_t \, \mathrm{d}t + \sqrt{2 \frac{\sigma^2}{m} \, K_t} \, \bullet \mathrm{d}W_t,
\qquad K_0=0,
\end{equation}
then the strictly negative drift coefficient of this equation would push the kinetic energy towards negative values,
without real physical meaning (even worse with complex values).
Finally, if one chooses the initially resting particle for the Itô case,
since $K_t=0$ is not an absorbing barrier for equation ~\eqref{kito} and the resulting drift is positive,
the particle is pushed towards a positive kinetic energy, which is the real physical effect of thermal fluctuations.

To analyze the general case of a not necessarily vanishing initial condition, we connect the process $K_t$ with a suitable squared Bessel process (BESQ).

\begin{proposition}\label{besq}
The unique process $K_t$ that solves the Itô equation \eqref{kito} subject to the fixed initial condition
$K_0=k_0:=\frac{1}{2}mv_0^2 \in \mathbb{R}_+$ is given by the explicit formula
\begin{equation*}
K_t=e^{-\frac{2\gamma}{m}t} \,\, \Theta\bigg(\frac{\sigma^2}{4\gamma}\big(e^{\frac{2\gamma}{m}t}-1\big)\bigg),
\end{equation*}
where $\big(\Theta(s)\big)_{s\geq 0}$ is a $BESQ^{\delta}_{k_0}$ process with dimension $\delta=1$ and initialized at $k_0$. Consequently, if the initial kinetic energy of the Langevin particle is positive, that is $k_0>0$, then it becomes zero in finite time almost surely.
\end{proposition}

\begin{proof}
The process $K_t$ determined by the Itô equation \eqref{kito} can also be rewritten in terms of a process widely used in finance and known as
the Cox-Ingersoll-Ross (CIR) process by simply renaming its parameters \cite{Jeanblanc}.
The unique solution to the CIR stochastic differential equation in terms of a squared Bessel process with dimension $\delta$ and initial condition $k_0$
(BESQ$^{\delta}_{k_0}$) is present in Chapter 6 of \cite{Jeanblanc}. Therefore the solution in the statement is found by the direct substitution of the
current model parameters in that solution. Finally, since all the parameters in equation ~\eqref{kito} are positive and $\delta=1$, the solution hits the
origin in finite time almost surely, again as it is shown in Chapter 6 of \cite{Jeanblanc}.
\end{proof}

\begin{remark}
For the unique solution of equation \eqref{kito}, the state $\{K_t=0\}$ is an instantaneously reflecting boundary \cite{Jeanblanc}. In particular,
this implies that the solution never takes negative values, which would be physically inconsistent for the current model. Moreover, the time set of the solution
becoming zero has null Lebesgue measure, as expected from the physical viewpoint too.
\end{remark}

\begin{remark}\label{zmft}
If the initial kinetic energy of the Langevin particle is positive, then it not only becomes zero in finite time almost surely: it also becomes zero
in mean finite time, which is a stronger condition that implies the former. The proof of this fact, along with an explicit representation formula for
the mean hitting time, can be found in \cite{cescudero}.
\end{remark}

Denote by $T_{0}^{k_0}$ the first hitting time of the origin by a Langevin particle initially possessing a positive kinetic energy $k_0$.
Since the trajectories of the process $K_t$ are continuous almost surely, then $K_t>0$ for all $t\in \big[0,T_{0}^{k_0}\big)$. Furthermore, as follows
from Proposition \ref{besq}, the interval $\big[0,T_{0}^{k_0}\big)$ is finite almost surely; note also that this interval, although of random duration,
can be tuned to arbitrarily small lengths by varying the initial condition, since $T_{0}^{k_0}$ collapses to zero in the limit of vanishing $k_0$.

Note that, for a given realization of the noise, equation \eqref{kito} can be solved explicitly (see either Proposition \ref{besq} or \cite{cescudero}),
so that $T_{0}^{k_0}$ becomes known.
While $t\in \big[0,T_{0}^{k_0}\big)$ equations \eqref{kito}, \eqref{strat}, and \eqref{khk} are equivalent; this follows from the smoothness of the
square root in this interval and Proposition \ref{proptrans}. Nevertheless, these equivalences fail at the moment the kinetic energy vanishes (what happens
in finite time almost surely) due to the lack of differentiability of the square root at the origin. Therefore, once the time $T_{0}^{k_0}$ has elapsed, the Stratonovich and H\"anggi-Klimontovich equations are affected by infinite multiplicity and non-existence of solutions respectively, as the situation is reminiscent to that of the null initial kinetic energy. Moreover, this phenomenon happens in a well-defined time scale, since
$\mathbb{E}(T_{0}^{k_0})<\infty$ by Remark \ref{zmft}. On the contrary, none of these difficulties affect the Itô equation, which possesses a unique
solution that is both global in time and physically meaningful.

For the sake of completeness, let us mention that a change of interpretation (from Stratonovich to Itô) that shifted the finite-time
blow-up of the solution to its global-in-time existence was proven in \cite{em} for a different physical model. In the present case,
the change of interpretation from Itô to H\"anggi-Klimontovich makes a global solution to cease to exist in finite time, but not to blow up.
Proposition \ref{besq} connected the solution to the Itô equation with a BESQ process of positive dimension,
characterized by an infinite lifetime \cite{pw}.
The shift of interpretation terminated the lifetime upon the collision of the process with the origin, an event that happens in finite time almost surely.
Interestingly, this behavior is characteristic of BESQ processes of negative dimension \cite{pw}. Overall, this illustrates the varied consequences that
a swap of interpretation might bring about. In connection to that, the fact that the solution ceases to exist in finite time implies that the long-time limit cannot be studied. This is important since the equipartition of energy should be fulfilled asymptotically in time. That is proven, via the fluctuation-dissipation relation, for the Itô equation \eqref{kito} in \cite{cescudero}. That reference also shows the difficulties that arise in such calculation for the Stratonovich equation \eqref{strat}. For the HK equation \eqref{khk} the analysis is simpler but more discouraging: the long-time limit does not even make sense, so the equipartition of energy is meaningless for that model. This is a direct consequence of the nonexistence of global in time solutions.

\subsection{Kinetic energy of a two Langevin particle system}

Consider a system of two independent Langevin particles. The velocities of the particles, denoted by $U_t$ and $V_t$, obey the stochastic differential equation system
\begin{equation}\label{system_langevin}
	\begin{split}
		m\,\mathrm{d}U_t &= - \gamma\, U_t\, \mathrm{d}t + \sigma\,\mathrm{d}B_t, \quad \quad U_0=u_0, \\
		m\, \mathrm{d}V_t &= - \gamma\, V_t\, \mathrm{d}t + \sigma\,\mathrm{d}W_t, \quad \quad V_0=v_0,
	\end{split}
\end{equation}
where $B_t$ and $W_t$ are independent Brownian motions both defined in $(\Omega,\mathcal{F},(\mathcal{F}_t)_{t\geq 0},\mathbb{P})$. Note that particles with independent velocities can be used as the starting point for the study of the properties of real gases \cite{lente2025direction}. Such as in the case of a single particle, the parameters $m, \gamma, \sigma >0$. From equation ~\eqref{system_langevin} it follows that both particles are immersed in the same medium and possess the same mass. The kinetic energy of the system, which in this case we denote by $\widetilde{K}_t$, is given by
\begin{equation*}
	\widetilde{K}_t=\frac{1}{2}\,m\big(U_t^2+V_t^2\big).
\end{equation*}
As done previously, we will obtain the stochastic differential equations that govern the evolution of $\widetilde{K}_t$ under the three different
interpretations of noise. To that end, consider the stochastic differential $\mathrm{d}\widetilde{K}_t=\frac{1}{2}\,m(\mathrm{d}U_t^2+\mathrm{d}V_t^2)$
along with Itô lemma and system \eqref{system_langevin}, to find out that
\begin{equation}\label{ket}
\mathrm{d}\widetilde{K}_t=\frac{\sigma^2}{m}\,\mathrm{d}t-2\frac{\gamma}{m}\widetilde{K}_t\,\mathrm{d}t+\sigma\big(U_t\,\mathrm{d}B_t+V_t\,\mathrm{d}W_t\big), \qquad \widetilde{K}_0=\frac{1}{2}\,m(u_0^2+v_0^2).
\end{equation}
To close an equation for $\widetilde{K}_t$ we need the following result.

\begin{proposition}
The stochastic process defined by
\begin{equation}\label{levyB}
	\widetilde{W}_t:=\int\limits_0^t\, \frac{U_s}{\sqrt{U_s^2+V_s^2}}\,\mathrm{d}B_s +\int\limits_0^t\, \frac{V_s}{\sqrt{U_s^2+V_s^2}}\,\mathrm{d}W_s
\end{equation}
is a Brownian motion.
\end{proposition}

\begin{proof}
To prove that $\widetilde{W}_t$ is a Brownian motion we will use the Lévy characterization theorem, see Theorem 8.4.2 in \cite{kuo}. The filtration $(\mathcal{F}_t)_{t\geq 0}$ and the probability measure $\mathbb{P}$, necessary for the application of the theorem, are the same that make up the filtered probability space $(\Omega,\mathcal{F},(\mathcal{F}_t)_{t\geq 0},\mathbb{P})$ in which $B_t$ and $W_t$ are defined. Using Theorems 4.6.1 and 4.6.2 of \cite{kuo}, since the integrands in Equation ~\eqref{levyB} have absolute value less than or equal to 1 almost surely, we know that the stochastic process $\widetilde{W}_t$ is a continuous martingale with respect to $(\mathcal{F}_t)_{t\geq 0}$ under $\mathbb{P}$.

Clearly, $\mathbb{P}(\widetilde{W}_0=0)=1$. In addition, the quadratic variation of $\widetilde{W}_t$ is given by
\begin{equation*}
  \langle\widetilde{W}_t\rangle =\int\limits_0^t\, \frac{U_s^2}{{U_s^2+V_s^2}}\,\mathrm{d}s +\int\limits_0^t\, \frac{V_s^2}{{U_s^2+V_s^2}}\,\mathrm{d}s=\int\limits_0^t 1\,\mathrm{d}s=t.
\end{equation*}
Thus, by Theorem 8.4.2 of ~\cite{kuo}, $\widetilde{W}_t$ is a Brownian motion.
\end{proof}

\begin{remark}
Using the language of stochastic differentials, the statement of this Proposition can be written in the form:
 \begin{equation}\label{dW}
	\sqrt{U_t^2+V_t^2}\,\mathrm{d}\widetilde{W}_t= U_t\,\mathrm{d}B_t + V_t\,\mathrm{d}W_t.
\end{equation}
\end{remark}

Now, if we use H\"anggi-Klimontovich calculus via Theorem \ref{Theorem_RCM}, we find
\begin{eqnarray}\nonumber
U_t \, \bullet \mathrm{d}B_t =  U_t \, \mathrm{d}B_t + \frac{\sigma}{m} \, \mathrm{d}t, \\ \nonumber
V_t \, \bullet \mathrm{d}W_t =  V_t \, \mathrm{d}W_t + \frac{\sigma}{m} \, \mathrm{d}t;
\end{eqnarray}
alternatively, using Stratonovich calculus yields
\begin{eqnarray}\nonumber
U_t \, \circ \mathrm{d}B_t =  U_t \, \mathrm{d}B_t + \frac{\sigma}{2m} \, \mathrm{d}t, \\ \nonumber
V_t \, \circ \mathrm{d}W_t =  V_t \, \mathrm{d}W_t + \frac{\sigma}{2m} \, \mathrm{d}t.
\end{eqnarray}
Our next step is to substitute Equation \eqref{dW} in \eqref{ket} and use the definition of $\widetilde{K}_t$ to arrive at the
Itô stochastic differential equation:
\begin{equation}\label{td_kito}
\mathrm{d}\widetilde{K}_t=\frac{\sigma^2}{m}\,\mathrm{d}t-2\frac{\gamma}{m}\widetilde{K}_t\,\mathrm{d}t+\sqrt{2\frac{\sigma^2}{m}\widetilde{K}_t}\,\mathrm{d}\widetilde{W}_t, \qquad \widetilde{K}_0=\frac{1}{2}\,m(u_0^2+v_0^2).
\end{equation}
If Stratonovich calculus is used the resulting equation is
\begin{equation}\label{td_stra}
\mathrm{d}\widetilde{K}_t=\frac{\sigma^2}{2m}\,\mathrm{d}t-2\frac{\gamma}{m}\widetilde{K}_t\,\mathrm{d}t+\sqrt{2\frac{\sigma^2}{m}\widetilde{K}_t}\circ\mathrm{d}\widetilde{W}_t, \qquad \widetilde{K}_0=\frac{1}{2}\,m(u_0^2+v_0^2).
\end{equation}
Finally, the H\"anggi-Klimontovich calculus yields
\begin{equation}\label{td_khk}
\mathrm{d}\widetilde{K}_t=-2\frac{\gamma}{m}\widetilde{K}_t\,\mathrm{d}t+\sqrt{2\frac{\sigma^2}{m}\widetilde{K}_t}\bullet\mathrm{d}\widetilde{W}_t, \qquad \widetilde{K}_0=\frac{1}{2}\,m(u_0^2+v_0^2).
\end{equation}

Analogously to the case of a single Langevin particle, we begin considering the two particle system initially at rest,
that is $\widetilde{K}_0=0$. Since $\widetilde{K}_t=0$ is an absorbing state for H\"anggi-Klimontovich equation \eqref{td_khk}, this means
that we are initializing the system at an absorbing boundary, which then becomes a valid solution for all times.
At the physical level, this translates into the two particle system remaining at rest forever or, in other words, the fluctuations from
the thermal bath disappear permanently. Of course, that is an absurd, and the overall situation is even worse, because this equation admits infinitely
many solutions, as will be shown at the end of this section. On the other hand, if we established the same null initial condition for the Itô and Stratonovich cases, respectively equations \eqref{td_kito} and \eqref{td_stra}, the null solution would not be allowed because $\widetilde{K}_t=0$ is not an absorbing state for either of these equations.
On the contrary, the state $\widetilde{K}_t=0$ is an instantaneously reflecting boundary for both equations, due to the presence of a positive inhomogeneous
term in them. Physically, this means that the system will progressively gain kinetic energy from rest due to the thermal fluctuations,
which is the actual physical picture.

If the system is initialized at some fixed positive kinetic energy $\kappa_0:=\frac{1}{2}\,m(u_0^2+v_0^2)$ rather than at rest,
then we can use the following proposition:

\begin{proposition}\label{besq_td}
The unique process $\widetilde{K}_t$ that solves the Itô equation \eqref{td_kito} subject to the initial condition
$\widetilde{K}_0=\kappa_0 \in \mathbb{R}_+$ is given by the explicit formula
\begin{equation*}
K_t=e^{-\frac{2\gamma}{m}t} \,\, \Xi\bigg(\frac{\sigma^2}{4\gamma}\big(e^{\frac{2\gamma}{m}t}-1\big)\bigg),
\end{equation*}
where $\big(\Xi(s)\big)_{s\geq 0}$ is a $BESQ^{\delta}_{\kappa_0}$ process with dimension $\delta=2$ and initialized at $\kappa_0$. Consequently, if the initial kinetic energy of the two Langevin particle system is positive, that is $\kappa_0>0$, then it stays strictly positive for all times
almost surely.
\end{proposition}

\begin{proof}
Based on the same transformation used in the proof of Proposition \ref{besq}, we can rewrite the process $\widetilde{K}_t$ in terms
of a $BESQ^{\delta}_{\kappa_0}$ process with dimension $\delta=2$ and initialized at $\kappa_0$ as specified in the statement.
Moreover, since all the parameters in equation ~\eqref{td_kito} are positive and $\delta=2$, the solution never hits the
origin in finite time almost surely, that is $\mathbb{P}(T^{\kappa_0}_0=\infty)=1$, where $T^{\kappa_0}_0$ is the first passage time to the origin
subject to the initial condition $\kappa_0$, see again Chapter 6 of \cite{Jeanblanc}.
\end{proof}

From Proposition \ref{besq_td} it becomes clear that the situation is not as serious as in the previous subsection. Since a two particle Langevin system
with an initial positive kinetic energy never possesses zero energy with probability one, the three stochastic calculi can be used interchangeably in
such a case. Still, if the initial condition is zero (or, for the same purpose, a random variable which distribution is atomized at the origin), then
the description in terms of the Itô or Stratonovich equation remains perfectly valid. However, the H\"anggi-Klimontovich equation \eqref{td_khk} presents
an uncountable number of solutions, such as the trivial one and the family:
\begin{equation*}
K_t=e^{-\frac{2\gamma}{m}(t-\tau(\omega))} \,\, \Xi_0\bigg(\frac{\sigma^2}{4\gamma}\big(e^{\frac{2\gamma}{m}(t-\tau(\omega))}-1\big)\bigg)
\,\mathlarger{\mathlarger{\mathbbm{1}}}_{t>\tau(\omega)},
\end{equation*}
for any $L^0(\Omega)$ and $\mathcal{F}_{0}-$measurable random variable $\tau(\omega)$ that is non-negative for almost all $\omega \in \Omega$, and
where $\big(\Xi_0(s)\big)_{s\geq 0}$ is a $BESQ^{\delta}_{0}$ process with dimension $\delta=2$ and initialized at $0$
(that for the case of the null initial condition; if the initial condition were a random variable with a positive probability of being zero, the expression
would need the corresponding modifications).
Of course, among all of these solutions (and the construction of more solutions is still possible, see for instance \cite{cescudero2}),
only one corresponds to the physical
reality: that with $\tau(\omega) \equiv 0$ almost surely. Since this is the unique solution of both the Itô and Stratonovich equations, it becomes clear
that both classical stochastic calculi are valid to study the kinetic energy of the two Langevin particle system, while
the H\"anggi-Klimontovich equation is not.

As a final remark, let us mention that the improved mathematical tractability observed in this subsection with respect to the previous one has a physical origin. A particle possesses zero kinetic energy if and only if its velocity is zero. That will happen recursively in the case of a fluctuating particle. However, a two-particle system will only possess zero kinetic energy if the velocities of both particles become zero simultaneously. That is an immediate consequence of the non-negativity along with the extensive character of the kinetic energy. Obviously, the second case will not happen by chance, and the system needs to be initialized at such a state. Similar observations were already employed in the mathematical analysis of the one-particle case in \cite{cescudero}. And they still provide a physically intuitive picture of the difference between the two-particle system and the single Langevin particle.

\subsection{Kinetic energy of the relativistic Brownian motion}

In the context of the Einstein theory of special relativity, a relativistic Brownian particle is a point particle embedded in a homogeneous thermal bath with constant temperature that, unlike its non-relativistic counterpart, possesses a speed $V_t$ that never exceeds the speed of light $c$ in absolute value;
for a more detailed description see references \cite{dunkelt,dunkel,dunkel2}.
In the two previous subsections we have studied the equations that govern the kinetic energy of one or two non-relativistic Langevin particles. In this subsection, we extend our study to the case of a single randomly dispersed relativistic particle. From now on we denote the rest mass of the relativistic particle by $M$ (and we assume it to be positive), the relativistic momentum by $P_t$, and the relativistic energy indifferently by $E_t$ or $P^0_t$.
They are related by the following defining formulas:
\begin{equation*}
    P_t:=M\,V_t\,\gamma(V_t), \qquad\qquad E_t:=c^2\,M\,\gamma(V_t) ,
\end{equation*}
where $\gamma(\cdot)$ is the Lorentz factor defined as
\begin{equation*}
    \gamma(v):=\frac{1}{\sqrt{1-\frac{v^2}{c^2}}}.
\end{equation*}
For simplicity, from now on we will adopt a system of natural units with $c=1$. By means of this simplification, we get
\begin{equation}\label{energy_relativistic}
    P_t=E_t\,V_t=M\,V_t\,\gamma(V_t), \qquad\qquad E_t \equiv P_t^0=(M^2+P_t^2)^{1/2}=M\,\gamma(V_t).
\end{equation}
Similarly to the Langevin approach to the non-relativistic case, the relativistic momentum process $P_t$ is assumed to obey an
Itô stochastic differential equation of the form \cite{dunkel2}
\begin{equation}\label{langevin_relativistic}
\mathrm{d}P_t= -\alpha(P_t)\,P_t\, \mathrm{d}t + [2D(P_t)]^{1/2}\,\mathrm{d}W_t, \quad \quad P_0=p,
\end{equation}
where $\alpha:\mathbb{R}\longrightarrow \mathbb{R}^{+}$ and  $D:\mathbb{R}\longrightarrow \mathbb{R}^{+}$ are the friction force and the amplitude of the thermal fluctuations respectively, while the initial condition $p \in \mathbb{R}$.
The specific form of the functions $\alpha(\cdot)$ and $D (\cdot)$ depends on the microscopic details of the interactions between the relativistic particle and the heat bath; for specific examples see Section 4.2.3 in \cite{dunkel2}. Herein we just assume them to be regular enough.

From the relation $P_t^0=(M^2+P_t^2)^{1/2}$, we define the new coefficients
\begin{equation*}\label{new_coef}
    \hat{\alpha}(P_t^0):=\alpha(P_t), \qquad\qquad \hat{D}(P_t^0):=D(P_t),
\end{equation*}
which we assume smooth enough so that any of the three types of stochastic calculi can be applied to them.
Equations for the relativistic energy process $P_t^0$ can then be derived from Equation \eqref{langevin_relativistic}, as has been done in \cite{dunkelt}.
The resulting Itô stochastic differential equation reads
\begin{equation}\label{kr_ito}
\mathrm{d}P_t^0= \Bigg\{-\hat{\alpha}(P_t^0)\,P_t^0\Bigg[1-\bigg(\frac{M}{P_t^0}\bigg)^2\Bigg]+\frac{\hat{D}(P_t^0)}{P_t^0}\bigg(\frac{M}{P_t^0}\bigg)^2\Bigg\}\,\mathrm{d}t + \Bigg\{2\hat{D}(P_t^0)\Bigg[1-\bigg(\frac{M}{P_t^0}\bigg)^2\Bigg]\Bigg\}^{1/2}\,\mathrm{d}W_t;
\end{equation}
while the Stratonovich one reads
\begin{equation}\label{kr_stratonovich}
\mathrm{d}P_t^0= \Bigg\{\bigg(\frac{\hat{D}^{\prime}(P_t^0)}{2}-\hat{\alpha}(P_t^0)\,P_t^0\bigg)\Bigg[1-\bigg(\frac{M}{P_t^0}\bigg)^2\Bigg]\Bigg\}\,\mathrm{d}t + \Bigg\{2\hat{D}(P_t^0)\Bigg[1-\bigg(\frac{M}{P_t^0}\bigg)^2\Bigg]\Bigg\}^{1/2}\circ\mathrm{d}W_t.
\end{equation}
Finally, the H\"anggi--Klimontovich stochastic differential equation is
\begin{eqnarray}\nonumber
\mathrm{d}P_t^0 &=&
\Bigg\{\Big(\hat{D}^{\prime}(P_t^0)-\hat{\alpha}(P_t^0)\,P_t^0\Big)\Bigg[1-\bigg(\frac{M}{P_t^0}\bigg)^2\Bigg]-\frac{\hat{D}(P_t^0)}{P_t^0}\bigg(\frac{M}{P_t^0}\bigg)^2\Bigg\}\,\mathrm{d}t
\\ \label{kr_hk}
&& + \Bigg\{2\hat{D}(P_t^0)\Bigg[1-\bigg(\frac{M}{P_t^0}\bigg)^2\Bigg]\Bigg\}^{1/2} \bullet \mathrm{d}W_t.
\end{eqnarray}
Instead of analyzing these equations in general, and in partial analogy to what has been done in the previous subsections,
we consider all three models subjected to the same initial condition:
\begin{equation}\label{ic_rc}
P_0^0 \equiv E_0=M;
\end{equation}
physically, this initial condition reflects the fact that the particle initially possesses a null relativistic momentum, i.e. $P_0=0$,
or in other words, it is at rest. It is direct to check that $P _t^0=M$ is an absorbing state for Equation \eqref{kr_stratonovich}; therefore, under
the considered initial condition, it is a solution too, not necessarily unique. Such a solution describes a state in which the particle remains at rest forever; something that of course is physically absurd in the presence of thermal fluctuations. For the H\"anggi-Klimontovich case, equation \eqref{kr_hk}, the same initial condition \eqref{ic_rc} leads to a negative drift simultaneously to a null diffusion, so the absorbing state transforms into an entrance
boundary. From the physical viewpoint, this is a nonsense too, since values of the relativistic energy lower than the rest mass are physically
unacceptable (not to speak of complex values).
Finally, the same initial condition \eqref{ic_rc} presents no problems for the Itô stochastic differential equation. Indeed,
$P_t^0=M$ is not an absorbing state for equation \eqref{kr_ito}, but rather an instantaneously reflecting state. At the physical level,
this means that a particle initially at rest will immediately gain relativistic energy due to the thermal fluctuations; that is nothing
but the real physical picture. In summary, we have shown that the pitfalls that were present in the case of the Langevin particle for
the Stratonovich and H\"anggi--Klimontovich interpretations, and absent for the Itô one, are replicated in the context of the relativistic Brownian motion.

Finally let us mention that there is, yet, one more advantage of the Itô interpretation in this model, which is actually tightly related to the previous one. The diffusion term in all the three equations \eqref{kr_ito}, \eqref{kr_stratonovich}, and \eqref{kr_hk} is the same: $\{2\hat{D}(P_t^0)[1-(M/P_t^0)^2]\}^{1/2}$. This term is not Lipschitz continuous but only Hölder$-1/2$ continuous at $P_t^0=M^+$ under the mild assumption $\hat{D}(M^+)>0$ (otherwise the thermal fluctuations would vanish when the particle is at rest). Note that the classical theory for existence and uniqueness of solution to SDEs demands Lipschitz regularity for the diffusion term \cite{kuo,oksendal}, but the Watanabe--Yamada theorem allows existence and uniqueness of solution to be proven for diffusion terms that are just as regular as Hölder$-1/2$, see \cite{wy}. However, the Watanabe--Yamada theorem applies only to Itô SDEs and not to Stratonovich ones \cite{ce,cescudero,cescudero2}. The fact that this theorem is not applicable to the H\"anggi--Klimontovich interpretation is a direct consequence of the previous subsection, which shows a counterexample. Therefore, the analysis of equation \eqref{kr_ito} is less problematic than that of equations \eqref{kr_stratonovich} and \eqref{kr_hk}, something that is, as already mentioned, tightly related to the previous discussion on the nature of the boundary at $P_t^0=M$, and overall shows the higher adaptability of the Itô integral to approach the dynamics of the kinetic energy of the relativistic Brownian motion.

\section{Further Extension of the Definition and Relation to the Backward Integral}\label{secextending}

Russo and Vallois introduced the backward integral (among others) in \cite{RV93} and they further analyzed it (them) in \cite{RV95,RV00,RV07};
it is defined as follows.

\begin{definition}
A stochastic process $\Upsilon_t,\ t \in [0,T] $, is said to be backward integrable (in the weak sense) with respect to a standard Brownian motion $W_t$,
if there exists another stochastic process $\mathcal{I}_t$ such that
\begin{equation}
\sup_{0 \le t \le T} \ \left\vert \int_0^t \Upsilon_s \, \dfrac{W_{s}-W_{s-\varepsilon}}{\varepsilon}\ \mathrm{d}s - \mathcal{I}_t\right\vert
\rightarrow 0\ ,\ \ \varepsilon \searrow 0
\label{backward_integral}
\end{equation}
in probability. If such a process exists, we denote
$$
\mathcal{I}_t:= \int_0^t \Upsilon_s \, \mathrm{d}^+ W_s,\ t\in[0,T],
$$
the backward integral of $\Upsilon_t$ with respect to $W_t$ over $[0,T]$.
\end{definition}

The following proposition shows a relation between the Russo--Vallois backward and the H\"anggi--Klimontovich integrals.

\begin{proposition}\label{propstepback}
Assume that $\Upsilon_t$ is a step process. Then it is backward integrable and the following convergence in probability
	\begin{equation}\nonumber
	\int_0^t \Upsilon_s \, \mathrm{d}^+ W_s = \lim_{\|\Delta_n\|\to 0}\sum_{j=1}^{n}\Upsilon_{t_j}(W_{t_j}-W_{t_{j-1}}),
	\end{equation}
holds true.
\end{proposition}

\begin{proof}
A step process is a stochastic process of the form:
$$
\Upsilon_t = \sum_{i=1}^{m} \zeta_{i-1} \mathlarger{\mathlarger{\mathbbm{1}}}_{(t_{i-1},t_i]}(t),
$$
where $\{\zeta_{i-1}\}_{i=1}^m$ is a collection of random variables. Then we have the trivial convergence:
\begin{eqnarray}\nonumber
\lim_{\|\Delta_n\|\to 0}\sum_{j=1}^{n}\Upsilon_{t_j}(W_{t_j}-W_{t_{j-1}}) &=& \sum_{i=1}^{m} \zeta_{i-1} (W_{t_i}-W_{t_{i-1}}) \\ \nonumber
&=& \sum_{i=1}^{m} \Upsilon_{t_i} (W_{t_i}-W_{t_{i-1}}).
\end{eqnarray}
On the other hand
\begin{eqnarray}\nonumber
\int_0^t \Upsilon_s \, \mathrm{d}^+ W_s &=& \lim_{\varepsilon \searrow 0} \int_0^t \Upsilon_s \, \dfrac{W_{s}-W_{s-\varepsilon}}{\varepsilon}\ \mathrm{d}s \\ \nonumber
&=& \lim_{\varepsilon \searrow 0} \sum_{i=1}^{m} \zeta_{i-1} \int_{t_{i-1}}^{t_i} \dfrac{W_{s}-W_{s-\varepsilon}}{\varepsilon}\ \mathrm{d}s \\ \nonumber
&=& \lim_{\varepsilon \searrow 0} \frac{1}{\varepsilon} \sum_{i=1}^{m} \Upsilon_{t_i}
\left( \int_{t_{i-1}}^{t_i} W_{s}\ \mathrm{d}s - \int_{t_{i-1}}^{t_i} W_{s-\varepsilon}\ \mathrm{d}s \right) \\ \nonumber
&=& \lim_{\varepsilon \searrow 0} \frac{1}{\varepsilon} \sum_{i=1}^{m} \Upsilon_{t_i}
\left( \int_{t_{i-1}}^{t_i} W_{s}\ \mathrm{d}s - \int_{t_{i-1}-\varepsilon}^{t_i-\varepsilon} W_{u}\ \mathrm{d}u \right) \\ \nonumber
&=& \lim_{\varepsilon \searrow 0} \frac{1}{\varepsilon} \sum_{i=1}^{m} \Upsilon_{t_i}
\left( \int_{t_{i}-\varepsilon}^{t_{i}} W_{s}\ \mathrm{d}s - \int_{t_{i-1}-\varepsilon}^{t_{i-1}} W_{s}\ \mathrm{d}s \right) \\ \nonumber
&=& \sum_{i=1}^{m} \Upsilon_{t_i} \lim_{\varepsilon \searrow 0}
\left( \frac{1}{\varepsilon} \int_{t_{i}-\varepsilon}^{t_{i}} W_{s}\ \mathrm{d}s - \frac{1}{\varepsilon} \int_{t_{i-1}-\varepsilon}^{t_{i-1}} W_{s}\ \mathrm{d}s \right)
\\ \nonumber &=& \sum_{i=1}^{m} \Upsilon_{t_i} (W_{t_i}-W_{t_{i-1}}),
\end{eqnarray}
where the last convergence happens almost surely, and hence in probability.
\end{proof}

This result does not mean that the H\"anggi--Klimontovich integral can just be considered as a kind of particular result of the backward one.
In the previous sections, we have introduced that integral respecting the way it is used in the physical literature, i.e. an integral that
gives rise to diffusion processes as solutions to its associated stochastic differential equations. In a certain sense, this means that
we have considered this integral as a perturbation of the It\^o integral, following the philosophy of the construction of the Stratonovich
integral in \cite{stratonovich}. In particular, this implies that the collection of potential integrands should be restricted to adapted
stochastic processes, contrary to what happens to the backward integral. The latter admits non-adapted integrands, what opens the possibility
to use it to build stochastic differential equations which solutions are not Markovian, not even adapted,
stochastic processes (which are, despite their reduced mathematical structure, of use in mathematical finance, see for instance
\cite{ee22,elizalde2025optimal,esc18}).

Nevertheless, it is useful to consider the backward integral in order to expand our definition of  H\"anggi--Klimontovich stochastic integration.

\begin{definition}\label{defweakhk}
Let $(\Omega,\mathcal{F},(\mathcal{F}_t)_{t\geq 0},\mathbb{P})$ be a completed filtered probability space in which a Brownian motion $(W_t)_{t\geq 0}$ is defined. A $\mathcal{F}_t$-adapted, almost surely square-integrable, stochastic process $\Upsilon_t,\ t \in [0,T] $, is said to be H\"anggi--Klimontovich integrable in the weak sense with respect to the standard Brownian motion $W_t$, if there exists another stochastic process $\mathcal{I}_t$ such that
\begin{equation}
\sup_{0 \le t \le T} \ \left\vert \int_0^t \Upsilon_s^{(n)} \, \dfrac{W_{s}-W_{s-\varepsilon}}{\varepsilon}\ \mathrm{d}s - \mathcal{I}_t\right\vert
\rightarrow 0\ ,\ \ \varepsilon \searrow 0, \ \ n \nearrow \infty,
\end{equation}
in probability, where $\{\Upsilon_t^{(n)}\}_{n=1}^\infty$ is a family of $\mathcal{F}_t$-adapted, almost surely square-integrable, step processes that approximate $\Upsilon_t$. If such a process exists, we denote
$$
\mathcal{I}_t:= \int_0^t \Upsilon_s \bullet \mathrm{d} W_s,\ t\in[0,T],
$$
the weak H\"anggi--Klimontovich integral of $\Upsilon_t$ with respect to $W_t$ over $[0,T]$.
\end{definition}

\begin{remark}\label{approxseq}
The family $\{\Upsilon_t^{(n)}\}_{n=1}^\infty$ always exists and approximates $\Upsilon_t$ in the sense that
$$
\int_0^t \vert \Upsilon_s^{(n)}  - \Upsilon_s \vert^2 \, \mathrm{d}s \rightarrow 0, \ \ n \nearrow \infty,
$$
in probability, see Lemma 5.3.1 in \cite{kuo}.
\end{remark}

\begin{remark}
By Proposition \ref{propstepback} and Remark \ref{approxseq} we know that the sequence of backward integrals generated by the approximating
sequence $\{\Upsilon_s^{(n)}\}_{n=1}^\infty$ (i.e. the vanishing $\varepsilon$ limit) in Definition \ref{defweakhk} always exists.
\end{remark}

This type of definition allows for more mathematical flexibility, as there are known advantages of introducing a theory of stochastic integration
by approximation \cite{kuo} or regularization \cite{RV07} rather than the classical Riemann-type discretization. We presume that this definition,
or a related one, could be of interest in the development of new mathematical physical results concerning the integral of H\"anggi and Klimontovich.
Or, perhaps, in its potential applications in other fields such as mathematical finance or engineering.

Finally, let us note other possible extensions of the H\"anggi--Klimontovich integral. First, to the anticipating setting by using,
as starting point, the extension of the Stratonovich integral proposed in \cite{jornet}. And second, to the spatially extended setting,
via its use in stochastic partial differential equations, employing, whether possible, the theoretical framework for the It\^o integral
exposed in \cite{dotti}.

\section{Conclusions}\label{conclusions}

The noise interpretation problem has received an enormous amount of attention along the decades in the physical literature. While classically,
the Stratonovich interpretation of noise seems to have been preferred within the realm of physics, during the recent years it is not difficult to
find works that propose the H\"anggi-Klimontovich interpretation as their favorite option. Whilst the question is undoubtedly interesting, there are
at least two remarks that, at least for us, are relevant in this topic. The first is that simplistic answers are most probably incorrect. The second
is that this question gets more blurred the less precise is the language employed to approach it. This is why we have conceived this work as a precise
mathematical analysis of the H\"anggi-Klimontovich interpretation of noise, but nevertheless we have tried to keep attention to the physical conclusions that can be extracted from it.

With that in mind, in this paper we have given a rigorous introduction to the H\"anggi-Klimontovich stochastic integral. Despite the popularity of this
noise interpretation in many recent physics articles, we are not aware of any prior precise mathematical introduction of it. Once the multidimensional
integral was introduced, a theory of H\"anggi-Klimontovich stochastic differential equations has been pushed forward. Under suitable assumptions, the
solutions to these equations are shown to be diffusion processes which probability distribution functions are shown to be governed by corresponding
Fokker-Planck equations. Perhaps not surprisingly given the importance of diffusion theory in physics, several properties of the solutions to
these Fokker-Planck equations have been highlighted as some of the advantages of the H\"anggi-Klimontovich interpretation. In this work we have provided
a demonstration of them along with some critical comments.

In connection with the previous paragraph, one may wonder why evaluating the integrands in the right endpoint of each subinterval in the Riemann-sum approximation should be highlighted in comparison to their evaluation at any other point. Such a general approach is highlighted in works like \cite{moreno2019conditional}. As we mentioned in the Introduction, given the relevance and mathematical characteristics of the Itô (left endpoint evaluation) and Stratonovich (midpoint evaluation) prescriptions, a symmetry argument only leaves the HK prescription to be explored. But of course, that should not be enough for some skeptical readers, who would demand a more tangible argument. Two such arguments can be found in section \ref{fpeab}. The first is the relative simplicity of the nonequilibrium potential, indeed, its simplest possible form is achieved under the HK interpretation, see Remark \ref{rem:noneqpot} in the present work and section~II in \cite{moreno2019conditional}. The second is the robustness of the deterministic behavior that is achieved under this interpretation, see Theorem \ref{linear_stability} and Remark \ref{rem:linest}. That is, the stable fixed points of the deterministic counterpart of a HK SDE correspond to metastable states of the stochastic system. This second fact is a consequence of the first, and it is only valid for the HK interpretation, as can be deduced from section \ref{fpeab}.

The previous-to-last chapter of this work has been devoted to the application of the H\"anggi-Klimontovich interpretation of noise to three different
examples of physical relevance: systems composed of one and two Langevin particles, and one relativistic Brownian particle. In the first and third cases,
both the Stratonovich and H\"anggi-Klimontovich interpretations gave physically inconsistent results, which were even more pathological for the latter (non-existence versus multiplicity of solutions), while the Itô interpretation was free from such problems. In the second case, both the Itô and Stratonovich interpretations were not problematic, but the H\"anggi-Klimontovich one was still affected by an infinite multiplicity of solutions.

Despite the advantages of the H\"anggi-Klimontovich interpretation of noise advertised in recent works, it also has clear mathematical disadvantages that manifest themselves in physically relevant models. Herein, we have highlighted some of those, which seem not to have been detected elsewhere in the literature. Perhaps paradoxically, what we have found is the absence of mathematical pathologies in the Itô interpretation, its presence in the Stratonovich one, and an even more severe manifestation of them in the H\"anggi-Klimontovich interpretation of noise. Although these facts contrast with some of the
lessons learnt in the physical literature along the decades \cite{Mannella,Kampen}, are in agreement with other mathematical physical developments \cite{ce,cescudero,cescudero2,em}. All of them show a higher mathematical flexibility of Itô integration that opposes the more rigid Stratonovich and H\"anggi-Klimontovich integrals. This is something that has to be taken into account in the construction of models affected by the interpretation of noise dilemma. After all, this dilemma was born as a consequence of the substitution of the formal manipulation of white noise by precise theories of stochastic integration.

\section*{Acknowledgements}

This work has been partially supported by the Government of Spain (Ministerio de Ciencia, Innovación y Universidades) and the European Union through Projects PID2024-158823NB-I00 and CPP2024-011557.

%This work has been partially supported by the Government of Spain (Ministerio de Ciencia e Innovaci\'on)
%and the European Union through Projects PID2021-125871NB-I00, CPP2021-008644/AEI/10.13039/501100011033/Uni\'on Europea NextGenerationEU/PRTR,
%and TED2021-131844B-I00/AEI/10.13039/501100011033/Uni\'on Europea NextGenerationEU/PRTR.

\vskip10mm
\noindent
{\footnotesize
Carlos Escudero\par\noindent
Departamento de Matem\'aticas Fundamentales\par\noindent
Universidad Nacional de Educaci\'on a Distancia\par\noindent
{\tt cescudero@mat.uned.es}\par\vskip1mm\noindent
}
\vskip2mm
\noindent
{\footnotesize
Helder Rojas\par\noindent
Departamento de Matem\'aticas Fundamentales\par\noindent
Universidad Nacional de Educaci\'on a Distancia\par\noindent
{\tt helder\_rojas@hotmail.com}\par\vskip1mm\noindent
}

\end{document}